\documentclass[twocolumn]{aastex7}

\newcommand{\mathbfit}[1]{\textbf{\textit{#1}}}

\usepackage{amsmath}

\begin{document}

\title{Modeling YSO Jets in 3D III: Dependence of Accretion and Jet Properties on Stellar Magnetospheric Field Strength and Rotation}

\author[0000-0003-2929-1502]{Yisheng Tu}
\affiliation{Department of Astronomy, University of Michigan, Ann Arbor, MI, 48109, USA}
\affiliation{Astronomy Department, University of Virginia, Charlottesville, VA 22904, USA}
\affiliation{Virginia Institute of Theoretical Astronomy, University of Virginia, Charlottesville, VA 22904, USA}
\email[show]{yitu@umich.edu}  

\author{Zhi-Yun Li}
\affiliation{Astronomy Department, University of Virginia, Charlottesville, VA 22904, USA}
\affiliation{Virginia Institute of Theoretical Astronomy, University of Virginia, Charlottesville, VA 22904, USA}
\email[]{zl4h@virginia.edu}  

\author{Zhaohuan Zhu}
\affiliation{Department of Physics and Astronomy, University of Nevada, Las Vegas, NV, 89154-4002, USA}
\affiliation{Nevada Center for Astrophysics, University of Nevada, Las Vegas, 4505 S. Maryland Pkwy, Las Vegas, NV, 89154, USA}
\email[]{zhaohuan.zhu@unlv.edu}  

\author{Kass Bell}
\affiliation{Astronomy Department, University of Virginia, Charlottesville, VA 22904, USA}
\affiliation{Virginia Institute of Theoretical Astronomy, University of Virginia, Charlottesville, VA 22904, USA}
\email[]{nnn8zm@virginia.edu}  

\begin{abstract}
    Observations of Young Stellar Objects (YSOs) systems reveal a wide diversity of jet properties, from well-collimated bipolar jets to uni-polar jets and systems with no detectable jet. Both prograde and counter-rotating jets are reported, raising questions about how jets are launched and how their properties relate to the underlying star-disk system. Using 3D non-ideal MHD simulations, we present a suite of models in which jet properties depend sensitively on stellar rotation and magnetic field strength. In all models, jets are launched from ``two-legged'' magnetic field lines anchored to both the star and the turbulent, magnetically elevated disk surface, with interactions at the disk surface crucial for mediating the magnetosphere-disk coupling. The axial jet and its surrounding disk wind form a characteristic ``spine-tower'' structure: the spine is the kinematically-dominated jet along open field lines threading the star, and the tower is the surrounding toroidal-field--dominated disk wind. The stability of this structure depends on the balance between the spine's stabilizing power and the tower's destabilizing power; if the tower dominates, the disk wind can choke the jet, producing asymmetric or no jets. This relationship allows an upper limit estimate on the toroidal magnetic field strength in the disk wind-launching region using observed outflow properties. Counter-rotating jets naturally appear in models, particularly with non-rotating stars, showing that the classical rotation-poloidal velocity relation does not reliably indicate the jet-launching radius. Instead, it could be used to trace the stellar rotation rate, offering a potential observational diagnostic of stellar spin.

\end{abstract}

\keywords{\uat{Young Stellar Objects}{1834} --- \uat{Circumstellar disks}{235} --- \uat{Jets}{870} --- \uat{Magnetohydrodynamics}{1964} ---\uat{Magnetic fields}{994} --- \uat{Stellar magnetospheres}{1610}}


\section{Introduction} 
\label{sec:introduction}
The stage of star formation during which a circumstellar disk is present is a crucial and relatively long-lived phase in the evolution of young stellar objects (YSOs). Circumstellar disks form as a natural outcome of the gravitational collapse of dense molecular cloud cores and serve as the birthplace of planetesimals and, ultimately, planets. A defining feature of this phase is the presence of outflows, which expel mass, energy, and angular momentum from the star-disk system back into the surrounding environment, thereby regulating both stellar growth and disk evolution \citep[as reviewed by, e.g., ][]{Frank2014, Bally2016, Pudritz2019, Ray2021}.

Outflows are ubiquitously observed in YSOs and display a remarkable diversity in morphology and kinematics. A commonly observed configuration is a multi-layered structure consisting of a fast, highly collimated outflow along the polar axis, typically referred to as the ``jet'', surrounded by a slower, wide-angle component often interpreted as a ``disk wind.'' \citep[e.g.,][]{Gudel2018, Lee2021, Vleugels2025} In many systems, the outflow appears approximately symmetric about the midplane \citep[e.g.][]{Lee2010, Lee2015, Qiu2019}; however, pronounced asymmetries are also observed, particularly in the jet, where one hemisphere hosts a strong jet while the opposite side shows a much weaker or even absent counterpart \citep[e.g., ][]{Podio2011, Hsieh2023, Assani2024, Liu2025, Dutta2026}. While some degree of asymmetry may arise from observational biases or projection effects, a growing number of cases are well established as intrinsic. In addition, jets are frequently observed to rotate, most often in the same sense as the disk \citep[][]{Coffey2011, Louvet2016, Lee2017, Lee2025}, yet counter-rotating jets have also been reported \citep[e.g.,][]{Coffey2004, Louvet2016}. This broad range of observed outflow properties poses a significant challenge for theoretical models, which must account for both the typical jet--disk-wind structure and the full diversity of asymmetry and rotation signatures.

A widely adopted framework for interpreting protostellar outflows envisions a two-component structure: a slower, wide-angle wind launched from the outer disk (the ``disk wind''), and a faster, more collimated jet originating from the inner disk/stellar vicinity region \citep[e.g., ][]{Lee2001, Matsakos2008, Fendt2009, Tabone2018}. In this picture, gas in the outer disk rotates more slowly and therefore produces a slower outflow, whereas gas in the inner disk rotates more rapidly and can potentially tap into a larger energy reservoir, with the star itself acting as an additional power source \citep[e.g.,][]{Romanova2005, Zanni2013}. How energy is extracted from the disk and/or the star to power the jet, however, has remained a long-standing problem. Given the high energetics of jets and extensive observational evidence for their ubiquity \citep[][]{Lee2001, Carrasco-Gonzalez2010, Goddi2017, Rodriguez-Kamenetzky2017, Rodriguez-Kamenetzky2025}, magnetic fields are widely believed to play a central role in the jet-launching process.

Classical theoretical models--including magneto-centrifugal launching models \citep[see, e.g.,][]{Blandford1982, Pudritz1983, Shu1994a, Shu1994b, Najita1994, Ostriker1995, Shu1995,  Shang2002, Cai2008} and magnetic tower/Poynting flux models \citep[][]{Shibata1986, Lynden-Bell1996, Lovelace2001, Lynden-Bell2003, Kato2004, Nakamura2007, Lovelace2009, Guan2014}--provide broad physical guidance for how rotational or magnetic energy can be converted into outflow kinetic energy. Nevertheless, the detailed mechanisms, and in particular how they operate in realistic star-disk systems, remain uncertain. Only in the past few decades have numerical simulations become sufficiently advanced to model the inner disk region from first principles with increasing levels of physical self-consistency \citep[e.g.,][]{Meliani2006, Stepanovs2014, Mizuno2022} and take into account the interaction between the stellar magnetosphere and the inner disk. \citet{Romanova2005} \citep[see also, e.g.][]{Ustyugova2006, Romanova2009, Lii2012, Romanova2014} put forward the ``propeller'' model, where conical-shaped outflows are launched by a rapidly spinning magnetized star. \citet{Zanni2009, Zanni2013} have shown that a similar conical-shaped outflow, which they term the ``magnetospheric ejection'', can also be launched by a slower rotating star, though such outflow may not be the typically observed jet.

Despite this progress, many existing models rely on the introduction of an \emph{ad hoc} or ``anomalous'' magnetic diffusivity in order to reach a quasi-steady state. In reality, stellar irradiation is expected to maintain a sufficiently high ionization fraction in the inner disk, placing the gas largely in the ideal-MHD regime, where the interaction between gas and magnetic fields can be significantly more dynamic and intermittent \citep[as reviewed by, e.g.,][]{Hartmann2016}. Additionally, most previous simulations have been carried out in two dimensions, where inherently three-dimensional magnetic processes--such as non-axisymmetric instabilities and turbulent field-line twisting--cannot be fully captured \citep[][]{Oishi2019}. Moreover, previous 3D simulations \citep[e.g., ][]{Lii2014, Zhu2024, Takasao2025, Zhu2025, Romanova2021} that incorporates a stellar magnetosphere mostly focus on the accretion processes associated with the disk-magnetosphere interaction, with limited focus on the outflow driven by this process.

To address these limitations, we have initiated a series of studies aimed at modeling YSO jets using fully three-dimensional, non-ideal MHD simulations. In the first paper of this series \citep[][]{Tu2025b}, we investigated a jet-launching system driven solely by the disk, neglecting both stellar rotation and star-disk magnetic interaction. We showed that such disk-only systems are prone to launching asymmetric or even one-sided outflows. In the second paper \citep[][]{Tu2026a}, we introduced a rotating star and its stellar magnetosphere, and demonstrated that the presence of a strong stellar magnetic field naturally promotes the formation of bipolar jets. These results suggest that a continuum of jet morphologies--from strongly one-sided to fully bipolar--exists between disk-dominated and magnetosphere-dominated regimes. Importantly, this connection implies that jet morphology and kinematics may encode valuable information about the underlying stellar and disk properties.

Building on \citet[][]{Tu2026a}, who presented and analyzed a single 3D simulation in detail, the present paper explores how robust this picture is across a broader parameter space. \citet[][]{Tu2026a} showed that the jet is launched along ``two-legged'' magnetic field lines that connect the star and the disk (a concept similar in spirit to those in, e.g., \citealp{Zanni2013} and \citealp{Lii2014}, but with different physical interpretation and consequences). Here, we investigate whether this mechanism persists when varying the relative polarity between the stellar and disk fields, the stellar rotation rate, and the stellar magnetic field strength. Understanding how these parameters modify the jet-launching process and the resulting outflow structure constitutes the central focus of this work.

This paper is organized as follows. In Section~\ref{sec:method}, we describe the numerical methods and model parameters. Section~\ref{sec:overview} provides an overview of the simulation results, highlighting key features in each model. In Section~\ref{sec:acc_outflow_depend_star}, we examine the relationship between jet properties and the underlying stellar parameters. Section~\ref{sec:discussion} discusses how our results compare to previous simulations and how the identified relationships can be used to infer star-disk system properties from observable outflow morphologies. Finally, we summarize our findings and conclude in Section~\ref{sec:conclusion}.

\section{Method}
\label{sec:method}
The inner disk region, where the stellar magnetosphere truncates the disk and launches an outflow, is governed by the following set of non-ideal MHD equations:
\begin{equation}
    \frac{\partial\rho}{\partial t} + \nabla\cdot(\rho\mathbfit{v}) = 0,
\end{equation}
\begin{equation}
    \rho\frac{\partial \mathbfit{v}}{\partial t} + \rho(\mathbfit{v}\cdot\nabla)\mathbfit{v} = -\nabla P + \frac{1}{c}\mathbfit{J}\times\mathbfit{B} - \rho\mathbfit{g},
    \label{equ:mhd momentum equation}
\end{equation}
\begin{equation}
    \frac{\partial \mathbfit{B}}{\partial t} = \nabla\times(\mathbfit{v}\times\mathbfit{B}) - \frac{4\pi}{c}\nabla\times(\eta_O\mathbfit{J}),
    \label{equ:mhd induction}
\end{equation}
where $\mathbfit{J} = (c/4\pi)\nabla\times\mathbfit{B}$ is the current density, and $\eta_O$ the Ohmic dissipation coefficient. Following \citet{Tu2026a}, ambipolar diffusion and the Hall effect are neglected in this study, as both are expected to play a relatively minor role in the near-ideal-MHD conditions of the star--disk interaction region. Other symbols have their usual meanings. 

We use the \texttt{Athena++} code \citep{Stone2020} to solve these equations in Cartesian coordinates with static mesh refinement (SMR). This work extends the models presented in \citet[][]{Zhu2024, Zhu2025, Tu2025b, Tu2026a}. The reference model is identical to the one analyzed in \citet[][]{Tu2026a}, while additional models are constructed by varying one simulation parameter from the reference model. Below, we summarize the key ingredients and parameters of the reference simulation, and the varying parameters of each model. We refer the readers to \citep[][]{Tu2025b, Tu2026a} for a full description of the reference model setup. 

The setup models the interaction between a stellar magnetosphere and a circumstellar disk. The stellar magnetosphere is represented by a dipolar magnetic field with magnetic field strength $B(R)\propto \bar{\mathbfit{m}} R^{-3}$, where $\bar{\mathbfit{m}}$ is the dipole moment. The magnetosphere is anchored to a star of mass $M_\star=1M_\odot$ and radius $r_\star \approx 0.015~\mathrm{au}$. The star itself is implemented as a dense sphere using the same treatment as in \citet[][]{Tu2026a}, where the hydrodynamic variables within $r_\mathrm{fix}=0.018~\mathrm{au}$ are reset to the initial values at each timestep to maintain stability. The circumstellar disk follows the same density and temperature structure as in \citet[][]{Tu2026a}, consisting of two zones: an active zone ($R \lesssim 0.13~\mathrm{au}$) where the gas is well coupled to the magnetic field (ideal MHD), and a dead zone ($R \gtrsim 0.13~\mathrm{au}$) where Ohmic dissipation dominates due to low ionization. For numerical stability and simplicity, the underlying temperature structure is reset to the initial condition at each time step, and the non-ideal MHD coefficient ($\eta_O$) is calculated using the Saha equation, including thermal ionization of alkali metals \citep[potassium, calcium, sodium, and magnesium, see][for more details]{Tu2025b}. We note that the jet-launching stellar magnetosphere-disk interaction region is almost entirely in the ideal MHD limit, so our result is expected to be minimally dependent on the computed diffusivity. The disk is initially threaded by a weak ordered poloidal magnetic field that threads the disk midplane vertically, identical across all models. 

To ensure numerical stability and a reasonable computational cost, the same density floor, as described in \citet[][]{Tu2026a}, is applied in all models, and we utilize the ``passive scalars'' in \texttt{Athena++} to distinguish between the added mass and the original mass. The latter is defined as the ``real gas'' in the simulation. We note that the added mass does not contribute significantly to the mass-loading of the jet.

Since this study focuses on the jets, and \citet[][]{Tu2026a} showed that the outflow is primarily launched along the boundary between the stellar magnetosphere and the disk surface, we vary the stellar magnetospheric parameters to test how jet properties depend on stellar conditions. In the reference model (named ``REF'' hereafter), the stellar rotation period is 12 days, and the surface dipole field strength is $B_\star \approx 2~\mathrm{kG}$, and its direction is aligned with the disk magnetic field.

To isolate specific physical effects, we conduct four additional models that differ from the reference case by only one parameter each.
\begin{enumerate}
    \item REV -- the stellar dipole is anti-aligned with the disk field, to test the effect of magnetic polarity.
    \item NRT -- the star is non-rotating, to separate the stellar and disk contributions to the jet power (since \citealp{Tu2026a} found the star contributes significantly).
    \item L3B -- the stellar surface magnetic field strength is reduced by factors of 3, to probe the effects of magnetic strength on star-disk coupling and jet launching.
    \item L6B -- similar to L3B, but lowering the stellar surface magnetic field strength by a factor of 6.
\end{enumerate}
All models share the same mesh refinement and cover the same time as the reference model (5.5 years).

\section{Result Overview}
\label{sec:overview}
In this section, we provide an overview of the simulations and describe the global properties of the system, laying the foundation for the detailed analysis and comparison presented in Section~\ref{sec:acc_outflow_depend_star}.

\begin{figure*}
    \centering
    \includegraphics[width=\textwidth]{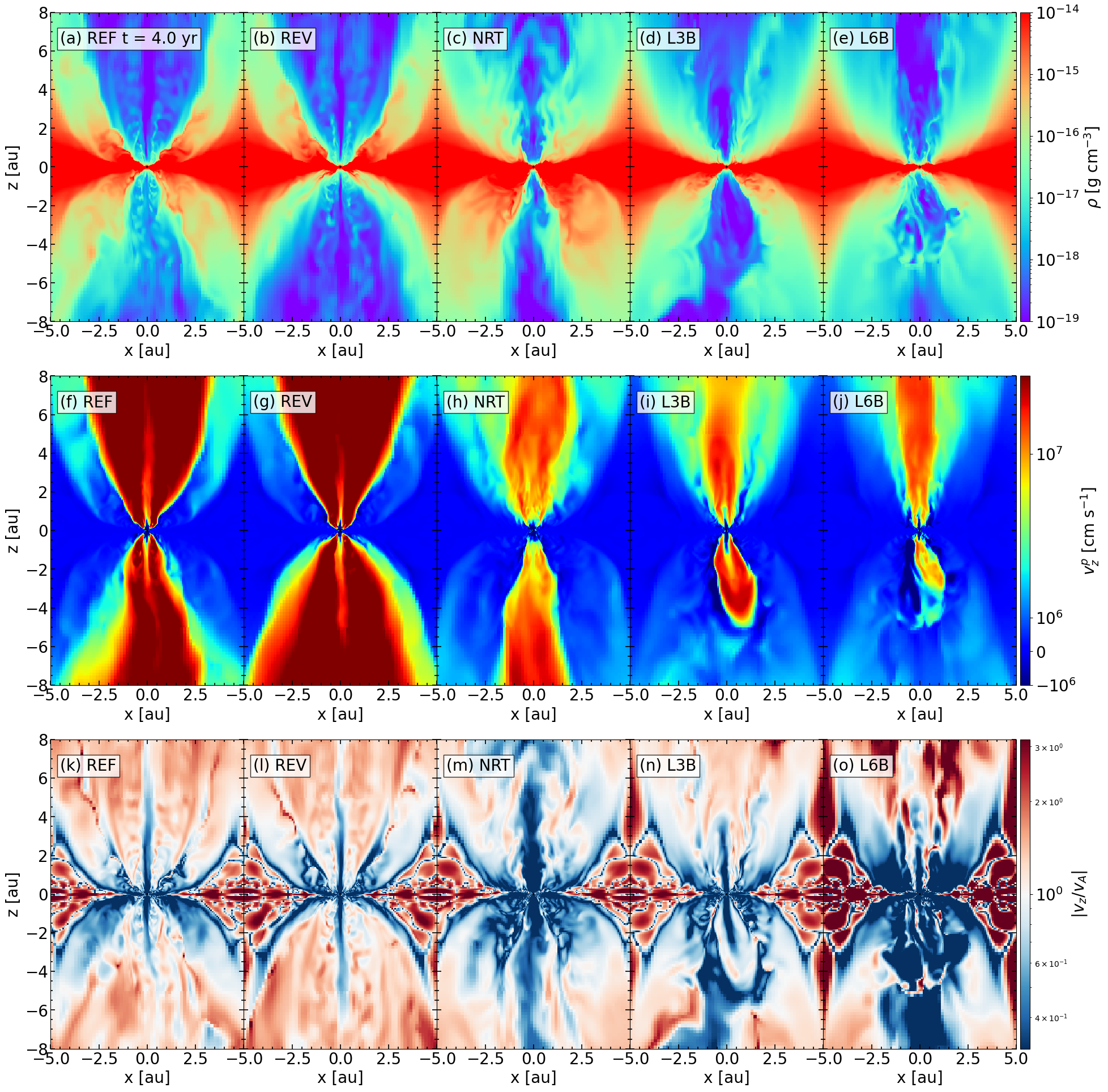}
    \caption{Overview of the disk, disk wind, and jet in each model at a representative time $t=4.0~\mathrm{yr}$. \textbf{Upper panels (panels a-e)} show the density on the meridian plane in each model, respectively (see section~\ref{sec:method} for a description of each model), exhibiting a high-density disk and a low-density outflow cavity in each model. \textbf{Middle panels (panels f-j)} show the corresponding projected $\hat{z}$-direction velocity ($v_z^p$, equation~\ref{equ:vzp}), where a positive value indicates an outflow away from the disk. The REF and REV models host strong, bipolar jets; the NRT, L3B, and L6B all host weaker and more intermittent jets. \textbf{Lower panels (panels k-o)} show the Alfv\'enic mach number ($|v_z/v_A|$) of the outflow, showing the jet is super-Alfv\'enic in the REF, REV, L3B, and L6B models, but remains sub-Alfv\'enic in the NRT model. An animated version can be found at \url{https://figshare.com/s/321aaeb17cf96f031902}. The animation is 22 seconds long and shows the evolution of all five models over the full simulation duration (5.5 yr). It highlights the persistent bipolar jet in the REF and REV models, as well as the more variable, often one-sided, or even absent jet in the NRT, L3B, and L6B models.}
    \label{fig:overview_2D}
\end{figure*}

Fig.~\ref{fig:overview_2D} shows a meridian slice through the simulation domain in each model at a representative time $t=4.0$~yr. The upper panels show the density distribution, and the middle panels show the projected $\hat{z}$-direction velocity, defined as
\begin{equation}
    v_z^p = v_z\frac{z}{|z|},
    \label{equ:vzp}
\end{equation}
where a positive value indicates an outflow away from the disk midplane. 

The outflow can be crudely divided into two components: the jet, defined as gas moving at $v_z^p>10^7~\mathrm{cm~s^{-1}}$, and a disk wind, defined as gas satisfying both moving at $10^6~\mathrm{cm~s^{-1}} < v_z^p \leq 10^7~\mathrm{cm~s^{-1}}$ and local $|B_z/B_\phi|<1$, to separate the disk wind from the thermally expanding disk atmosphere (see figure~\ref{fig:energy_flux_z}). The jet in all models roughly coincides with a low-density polar cavity, where the density $\lesssim 10^{-18}~\mathrm{g~cm^{-3}}$; the disk wind surrounds the jet in all models, with a higher density of $\lesssim 10^ {-16}~\mathrm{g~cm^{-3}}$.

Overall, the REF and REV models produce strong, clearly bipolar jets with similar morphology. The jet in the NRT model tends to be bipolar, although it can sometimes be unipolar at high altitudes, and the jet speed is lower, indicating a weaker jet. The jet in the L3B and L6B models varies over time between bipolar, unipolar, and sometimes no jet at large distances, and the jet speed is generally slower than that in the REF and REV models. 

To further illustrate the jet kinematics, the lower panels of Figure~\ref{fig:overview_2D} show the Alfv\'enic Mach number, defined as
\begin{equation}
\mathcal{M} \equiv \Big|\frac{v_z}{v_A}\Big| = \Big|\frac{v_z\sqrt{4\pi\rho}}{B}\Big|,
\end{equation}
where $B$ is the total magnetic field strength. In the REF and REV models, the jet transitions from sub-Alfv\'enic to mostly super-Alfv\'enic at relatively low altitudes ($\sim2~\mathrm{au}$). By the time the flow reaches the simulation boundary ($z=\pm20~\mathrm{au}$), the jet is entirely super-Alfv\'enic.

In contrast, in the NRT model, where the disk is the sole energy source, the jet remains sub-Alfv\'enic even at large heights. This is primarily because, although the magnetic flux is comparable to that in the REF and REV models, both the mass-loss rate and outflow velocity are lower. A more detailed analysis of the magnetic flux budget is presented in Sections~\ref{sec:mag_flux_evol} and \ref{sec:opened_stellar_flux}.

The L3B and L6B models exhibit variable jets. When the jet is fully developed, it becomes super-Alfv\'enic, similar to the REF model; however, during phases when the jet is choked, the outflow remains sub-Alfv\'enic due to its significantly reduced velocity.

Before analyzing the physical drivers of these morphological differences, we first present the global properties of the system in Sections~\ref{sec:acc_outflow_rates}, \ref{sec:mag_flux_evol}, and \ref{sec:jet_origin}. A detailed investigation of the jet-launching mechanism and its model dependence follows in Section~\ref{sec:acc_outflow_depend_star}.

\subsection{Accretion and outflow rates}
\label{sec:acc_outflow_rates}
\begin{figure*}
    \centering
    \includegraphics[width=\textwidth]{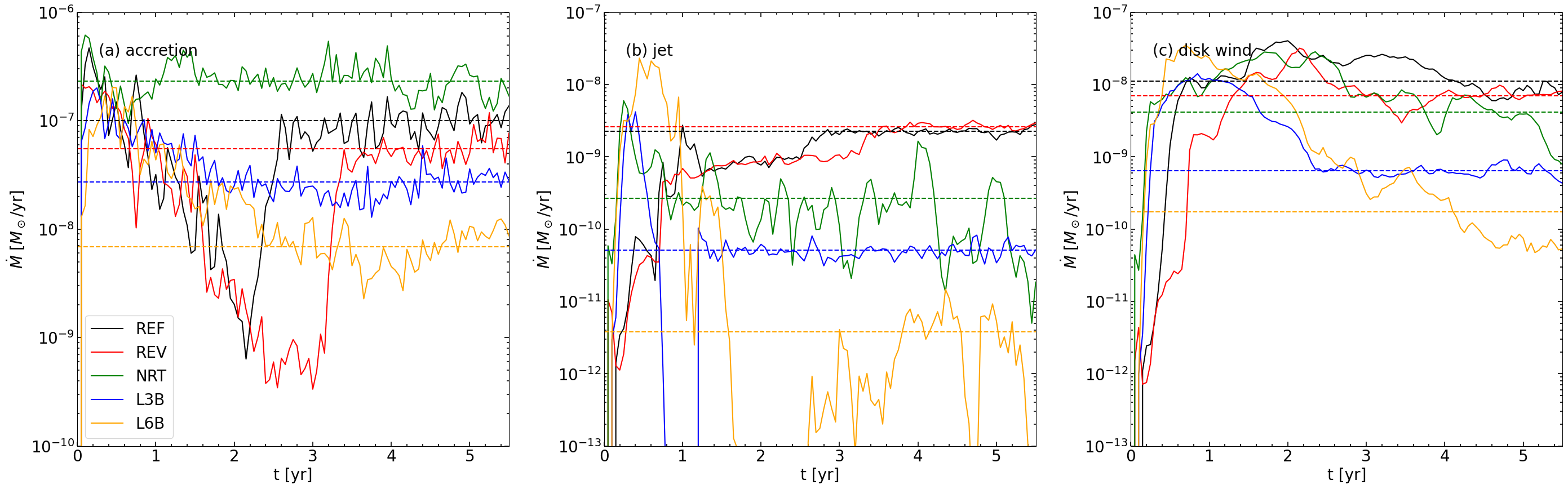}
    \caption{Overview of accretion and outflow properties in each model. \textbf{Panel (a)} shows the accretion rate in each model, measured at $r=0.02~\mathrm{au}$ and only account for the mass of the ``real gas'' (see section~\ref{sec:method}). The values are multiplied by $-1$, so a positive value in this panel is accretion; \textbf{panel (b) and (c)} show the jet and disk wind outflow rate in each model, respectively. Both are measured at $z=\pm5~\mathrm{au}$ and include only the ``real'' mass. The horizontal dashed lines in each panel show the averaged accretion/outflow rates between 3.5~yr and 5.5~yr, where the accretion/outflow reaches a quasi-steady value.}
    \label{fig:overview_acc_out}
\end{figure*}

We first present the accretion and outflow rates in the simulation. The accretion rate in each model is shown in Figure~\ref{fig:overview_acc_out}(a), measured at $r=0.02~\mathrm{au}$ and only includes the ``real gas'' (see section~\ref{sec:method}). 

In all models with stellar rotation (i.e., REF, REV, L3B, and L6B), the accretion rate drops in the first $\sim 2$ to $3$ yrs, and settles to a quasi-constant value after. This decrease in accretion rate is caused by the rotating star injecting angular momentum into the disk, thereby decreasing the accretion rate (more details in section~\ref{sec:stellar_rotation}). After the initial $\sim 2$ to $3$ years, the system relaxes into a quasi-equilibrium, so the accretion rate becomes more steady thereafter. A horizontal dashed line is presented for each model to show the averaged accretion in the quasi-equilibrium state between 3.5 yr and 5.5 yr.

The accretion rate in the NRT, no-rotating star, model is the highest across all models, holding at about $2.2\times10^{-7}~M_\odot/\mathrm{yr}$. The REF model seconds around $1\times10^{-7}~M_\odot/\mathrm{yr}$, and the REV model follows at around $5.3\times10^{-8}~M_\odot/\mathrm{yr}$. These three models (NRT, REF, and REV) share the same initial stellar magnetosphere field strength. They only differ in the relative field orientation (REV) and stellar rotation rate (NRT). The models with lower initial stellar magnetosphere field strength have lower accretion rates, staying around $2.8\times 10^{-8}~M_\odot/\mathrm{yr}$ in the L3B model and $7\times10^{-9}~M_\odot/\mathrm{yr}$ in the L6B model.

The outflow rates of the jet and the disk wind are presented in figure~\ref{fig:overview_acc_out}(b) and (c) respectively, including only the ``real gas'' and measured at $z=\pm10~\mathrm{au}$. Since the outflow is divided into the jet ($v_z^p>10^7~\mathrm{cm~s^{-1}}$) and the disk wind ($10^6~\mathrm{cm~s^{-1}} < v_z^p < 10^7~\mathrm{cm~s^{-1}}$ and local $|B_z/B_\phi|<1$), we measure their outflow rates separately. Similar to the accretion rate, the averaged outflow rate between 3.5~year and 5.5~year is presented by a horizontal dashed line. 

The jet outflow rates in the REF and REV models are comparable, both reaching $\approx 2\times10^{-9}~M_\odot~\mathrm{yr^{-1}}$. These values are roughly $50\times$ and $25\times$ smaller than the accretion rates in the REF and REV models, respectively. The disk-wind mass-loss rates in both models are also similar, at $\sim10^{-8}~M_\odot~\mathrm{yr^{-1}}$, corresponding to $10\sim20\%$ of the accretion rate.

In the NRT model, both the jet and disk-wind outflow rates are significantly lower--by factors of $\sim10$ and $\sim5$, respectively, relative to the REF and REV models. This reduction reflects the absence of stellar rotation as an energy source for launching and sustaining the outflow. We will quantify the stellar contribution to the energy budget in Section~\ref{sec:stellar_rotation}.

The L3B and L6B models, which have weaker stellar magnetic fields, exhibit reduced jet and disk-wind mass-loss rates even though the star is rotating. This indicates that the efficiency of magnetic energy transfer from the star to the outflow declines sharply as the stellar field weakens. A quantitative assessment of this trend is presented also in Section~\ref{sec:stellar_rotation}.

Because both accretion and outflow rates depend sensitively on the magnetic field strength and topology, we now examine the magnetic flux evolution in the following subsection.

\subsection{Magnetic flux evolution}
\label{sec:mag_flux_evol}
The magnetic field can be roughly divided into 4 zones during the later, quasi-static, stage of the simulations, as outlined in \citet[][see their Figure~2]{Tu2026a}. One of the four zones is the permanently closed stellar magnetosphere; the other three zones are results of the partial opening of the initial stellar magnetosphere: a stellar field threading the star; a disk field threading the disk; and a highly dynamic region between the first two. 

\begin{figure*}
    \centering
    \includegraphics[width=\linewidth]{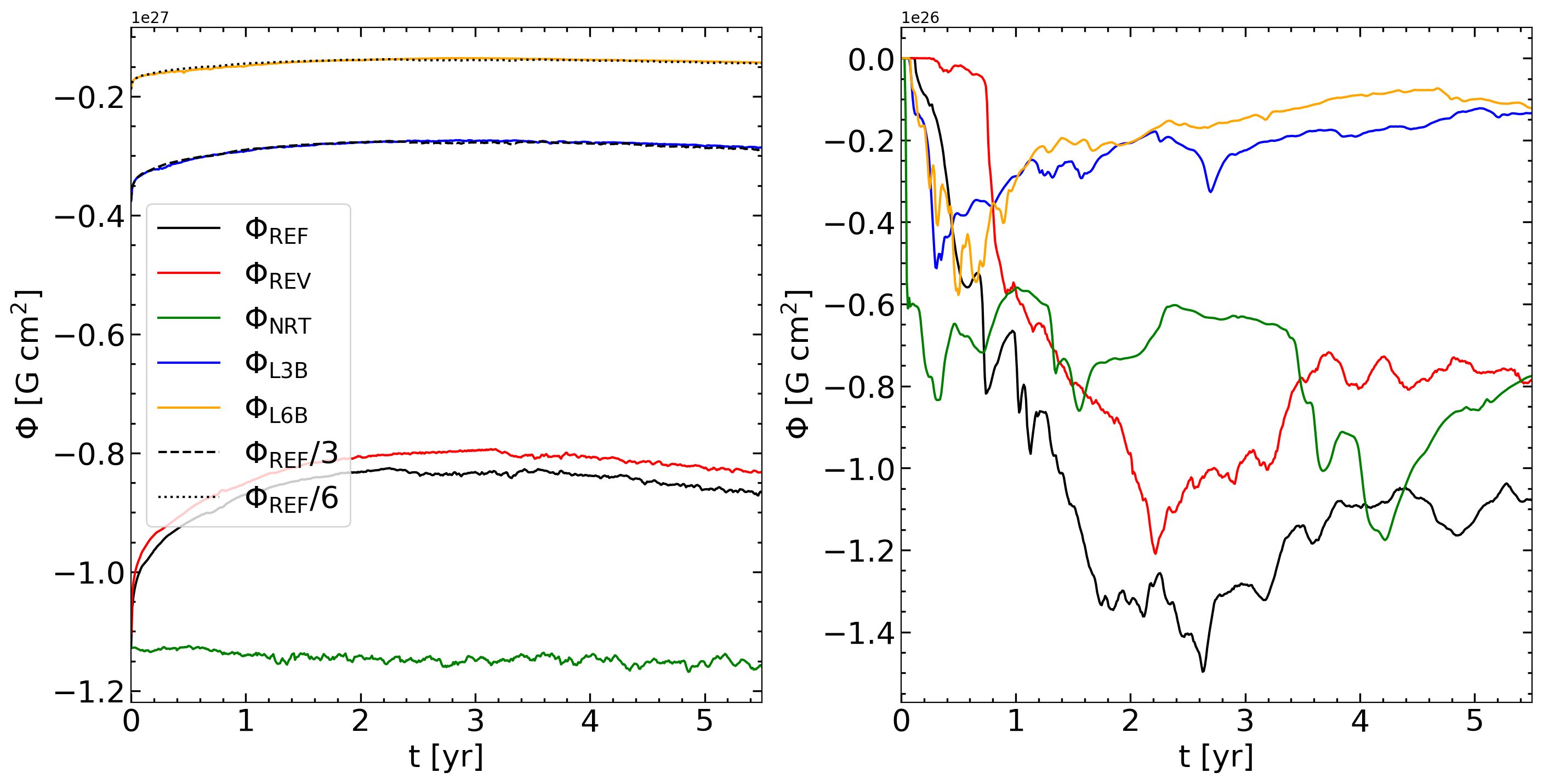}
    \caption{Magnetic flux evolution in each model. \textbf{Panel (a)} shows the amount of closed dipolar magnetosphere flux in each model (see section~\ref{sec:mag_flux_evol}), with the expected flux in the L3B and L6B models scaled from the REF model. \textbf{Panel (b)} shows the amount of opened stellar magnetosphere flux (see section~\ref{sec:mag_flux_evol}), measured at $z=\pm5~\mathrm{au}$ and averaged over both hemispheres. }
    \label{fig:models_fluxes}
\end{figure*}

Although the permanently closed stellar magnetosphere does not directly participate in launching the outflow, it serves as a reservoir of magnetic flux that contributes to the total available flux in the system. To quantify the evolution of the stellar magnetosphere, Figure~\ref{fig:models_fluxes}(a) shows the total downward-directed flux through the disk midplane\footnote{For the magnetic flux calculation, \citet[][]{Tu2026a} also did a similar calculation in their Figure~3. They defined the flux as the maximum magnitude of the azimuthally integrated downward flux at each radius, whereas here we use the total downward flux at a given height. The former tends to underestimate the flux, while the latter may overestimate it, but both measures produce generally consistent results.}. Because the dipolar magnetosphere is represented in the simulation by closed magnetic loops, some flux gain or loss due to reconnection is expected, especially in models with a rotating star.

In the non-rotating NRT model, the magnetospheric flux remains nearly constant over time ($\sim 1.15\times10^{27}~\mathrm{G~cm^2}$). In the REF and REV models—both initialized with the same stellar field strength as the NRT model--roughly one quarter of the initial flux is lost through reconnection (i.e., $\Phi\approx 8.5\times10^{26}~\mathrm{G~cm^{2}}$. Comparable fractional losses are also seen in the L3B and L6B models, and their absolute flux levels ($2.9\times10^{26}$ and $1.5\times10^{26}~\mathrm{G~cm^{2}}$, respectively) follow the expected scaling with the reduced initial stellar field strengths.

More relevant to driving both the outflow and accretion is the amount of opened stellar-magnetosphere flux, which sets the total magnetic flux available in the inner disk and in the interaction region where the jet is launched (see Section \ref{sec:jet_origin}). Because all downward-directed magnetic field at high altitudes in each model comes from the partial opening of the stellar magnetosphere, we estimate the opened flux by measuring the vertically averaged downward flux at $\pm 2~\mathrm{au}$, as shown in Figure~\ref{fig:models_fluxes}(b). The relatively low altitude of this measurement is chosen because the jet (hence the downward-direction flux in the polar region) is better presented at this altitude; at a larger altitude, the jet can be choked by the turbulent disk wind (Figure~\ref{fig:overview_2D}), where the field geometry is more disordered.

Despite the roughly 25\% difference in the closed-magnetosphere flux among the REF, REV, and NRT models, the amount of opened flux is essentially the same in all three, approximately $10^{26}~\mathrm{G~cm^2}$. In contrast, the L3B and L6B models exhibit an opened flux of only $0.2\times10^{26}~\mathrm{G~cm^2}$ and $0.1\times10^{26}~\mathrm{G~cm^2}$ respectively, smaller than expected from a simple scaling of the REF model. We discuss the expectation and apparent discrepancy in greater detail in Section~\ref{sec:opened_stellar_flux}.

The evolution of the magnetic flux is what ultimately gives rise to the various outflow properties observed across the different models. Before analyzing the physical origins of these differences, we first identify the regions from which each outflow component is launched. This establishes the necessary context and guides the detailed examination of outflow and accretion presented in Section~\ref{sec:acc_outflow_depend_star}.
\subsection{Origin of the jet}
\label{sec:jet_origin}
\citet[][]{Tu2026a} showed that the jet is launched along ``two-legged'' magnetic field lines, with one footpoint anchored on the stellar surface and the other embedded in the disk. These field lines drive the jet through a ``load-fire-reload'' cycle: differential rotation between the star and disk rapidly builds up a strong toroidal field during the load phase; this toroidal field then powers an outflow during the fire phase, opening the field lines; subsequent magnetic reconnection restores the original geometry and recreates new two-legged field lines, allowing the cycle to repeat. Because this process operates on very short timescales and proceeds simultaneously across different azimuths and radii, the resulting jet appears continuous at large distances from the system. The concept is similar to the ``field inflation'' concept in, e.g., \citet{Lynden-Bell1994, Lovelace1995, Zanni2013}, but the cycles repeat on a much shorter time-scale.

An important implication of this mechanism is that the jet mass originates primarily from the disk surface, while its energy is supplied by both the star and the disk. We quantify the associated energy budget in Section~\ref{sec:stellar_rotation}, and in this section we demonstrate that the same basic mechanism operates across all models, emphasizing in particular the disk origin of the jet mass.

\begin{figure*}
    \centering
    \includegraphics[width=\linewidth]{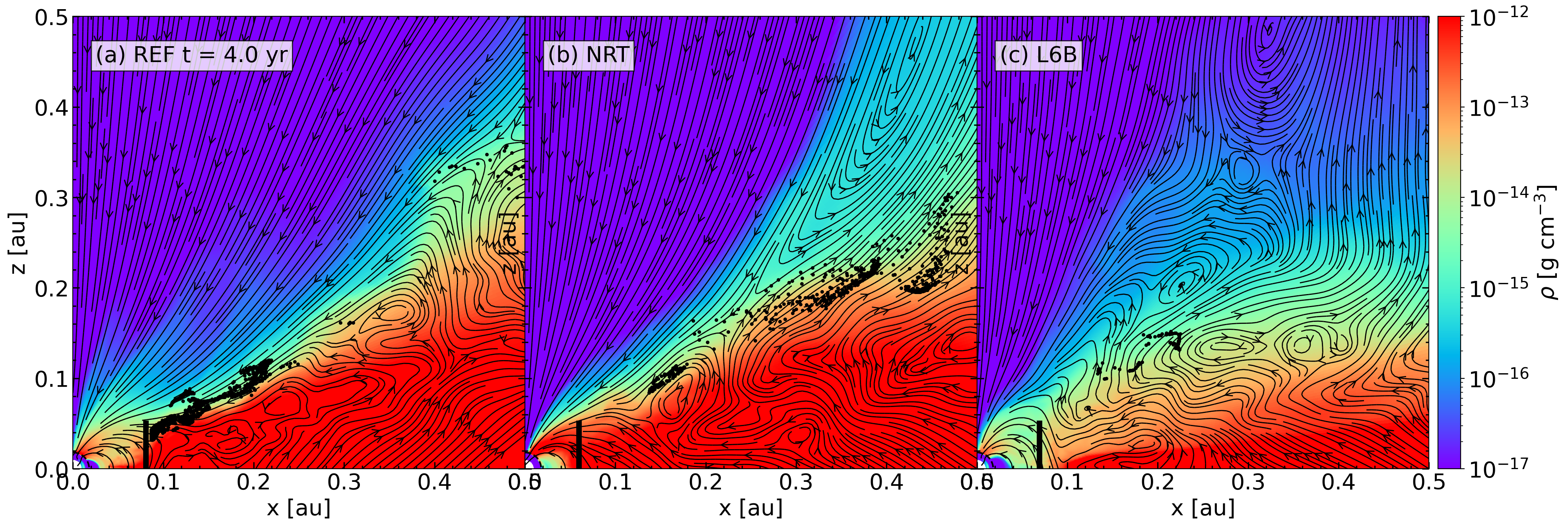}
    \caption{Root of the jet projected onto the meridian plane. Each dot marks the end point of a streamline traced backwards from the jet using the instantaneous velocity profile (section~\ref{sec:jet_origin}). The background is azimuthal-averaged ``real'' density as a reference, overplotted with azimuthal-averaged magnetic field lines. The thick vertical black line is located at the averaged truncation radius in each model (see section~\ref{sec:trunc_radius}), marking an approximate location for the truncation radius. These three panels show the REF, NRT, and L6B models, respectively.}
    \label{fig:mass_loading}
\end{figure*}

To establish that the jet mass is loaded from the disk surface, we trace jet velocity streamlines backward to determine whether they originate from the disk. The streamline seeds are placed within the jet (defined as regions with $v_z^p > 10^7~\mathrm{cm~s^{-1}}$) at $z = 5~\mathrm{au}$. Each streamline is then traced backward until either (1) the gas enters a region with $\beta_K > 1$ and $\rho > 10^{-14}~\mathrm{g~cm^{-3}}$, indicating disk material, or (2) the integration fails due to numerical constraints\footnote{Because the system is highly dynamic and time-dependent, streamlines traced at a single snapshot may occasionally enter closed or extremely long loops. Such streamlines are excluded from our analysis.}. Figure~\ref{fig:mass_loading} shows the terminal points of these back-traced streamlines, projected onto the meridional plane. The background displays the azimuthally averaged density and magnetic field lines. The three panels show the REF, NRT, and L6B models, with the remaining models and times exhibiting similar behavior.

All streamline origins lie on the disk surface. Notably, every origin point is located near or outside the truncation radius, indicated by the thick black vertical line for each model (see section~\ref{sec:trunc_radius}). This spatial distribution demonstrates that the outflow is not fed by the star nor by material flowing along the rigid stellar magnetospheric funnel. Instead, the jet is loaded directly from gas on the disk surface and then accelerated magnetically by the two-legged field lines. 

\section{Dependence of Disk and Outflow Properties on Stellar Magnetospheric Field Strength and Rotation}
\label{sec:acc_outflow_depend_star}
\subsection{Disk truncation radius}
\label{sec:trunc_radius}
One of the most important quantities characterizing the disk-magnetospheric interaction is the truncation radius: the location where the strong stellar magnetic field disrupts and truncates the accretion disk. 

\begin{figure}
    \centering
    \includegraphics[width=\linewidth]{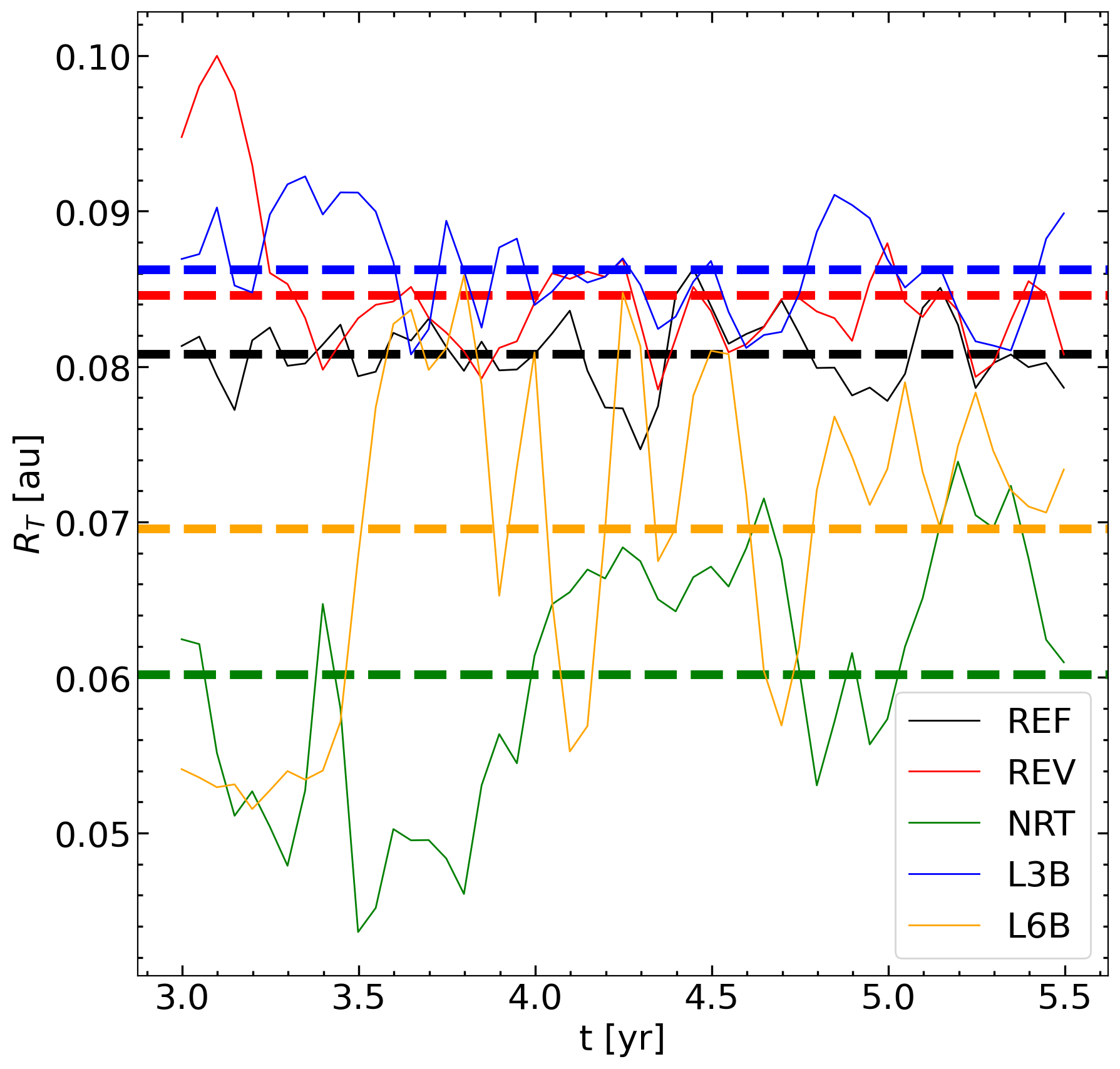}
    \caption{The truncation radius, defined as the innermost radius where $\beta_K=1$ (section~\ref{sec:trunc_radius}), in each model as a function of time. The horizontal thick dashed line shows the averaged truncation radius in each model.}
    \label{fig:truncation_radius}
\end{figure}

To measure the truncation radius in each simulation, we follow \citet{Zhu2024}, \citet{Takasao2025}, and \citet[][]{Tu2026a}, defining the truncation radius $R_T$ as the innermost radius at which $\beta_K = 1$. The evolution of $R_T$ in the later stages of each model is shown in Figure~\ref{fig:truncation_radius}. In the REF, REV, L3B models, $R_T$ remains close to $0.085~\mathrm{au}$; in the L6B models, $R_T$ is smaller, around $0.07~\mathrm{au}$, and in the NRT model, $R_T$ is the smallest, around $0.06~\mathrm{au}$.

While the $R_T$ in the L6B model is expected to be smaller due to the weaker magnetic field, the NRT model, whose magnetic field strength is the same as the REF model, has the smallest truncation radius, which may appear surprising at first glance, since one might expect the truncation radius to depend directly on the stellar magnetosphere strength. However, this expectation is not borne out when compared with the analytic estimate of \citet{Bessolaz2008}: {\footnote{We caution the reader, however, that direct application of $R_\mathrm{T, ana}$ in our model require some special caution, as we also include a disk field in the simulation. Although relatively weak compared to the stellar magnetosphere, it could change the effective stellar dipole moment $\bar{m}$ and mass accretion rate $\dot{M}$ by altering the global magnetic flux budget. As a consequence, the value of $R_\mathrm{T, ana}$, calculated through initial prescribed $\bar{m}$ and averaged $\dot{M}$ (figure~\ref{fig:overview_acc_out}), should be taken as a rough estimate of the expected truncation radius in the simulation, and we have verified this estimation is reasonably close to the values measured through the $\beta_K$ definition shown in figure~\ref{fig:truncation_radius}.}}
\begin{equation}
    R_{T,\mathrm{ana}} = R_\star \Big(\frac{B_\star^4 R_\star^5}{2 G M_\star \dot{M}^2}\Big)^{1/7} = \Big( \frac{\bar{m}^4}{2 G M_\star \dot{M}^2} \Big)^{1/7},
    \label{equ:Rt_ana}
\end{equation}
where $\bar{m}$ is the dipole moment \citep[see equation~8 in][]{Tu2026a}, which is linearly proportional to the stellar magnetic field strength, and $M_\star$ and $\dot{M}$ are the stellar mass and accretion rate, respectively. For a fixed stellar mass, the scaling reduces to $R_T \propto (\bar{m}^4 / \dot{M}^2)^{1/7}$.

By definition, the accretion rate is $\dot{M} = \rho A v_R$, where $\rho$ and $A$ are the density and flow cross-section; both are similar across all models. Thus the differences in $\dot{M}$ are set primarily by $v_R$, the radial inflow velocity. In the inner disk's magnetically active region, the accretion velocity is controlled by magnetic braking, which depends on the magnetic field strength. \citet{Tu2025a} showed that $v_R \propto B_\phi B_z$. In our simulations, both $B_\phi$ and $B_z$ scale linearly with the stellar magnetosphere strength (i.e., with $\bar{m}$), and all models share a similar density $\rho$, so that $\rho v_R \propto B^2 \propto \bar{m}^2$.

Substituting this into the analytic scaling yields
\begin{equation}
    R_T \propto \Big(\frac{\bar{m}^4}{v_R^2} \Big)^{1/7}\propto\Big(\frac{\bar{m}^4}{\bar{m}^4} \Big)^{1/7}\propto 1,
\end{equation}
showing that the truncation radius is expected to be approximately independent of the stellar magnetic field strength--consistent with what we find in the simulations. Physically, a decrease in the stellar magnetospheric field strength, $\bar{m}$, reduces the magnetic flux threading the disk and therefore weakens the magnetic torque that drives accretion, leading to a lower mass accretion rate, $\dot{M}$. In equation~\ref{equ:Rt_ana}, the dependences on $\bar{m}$ and $\dot{M}$ largely compensate for each other: as $\bar{m}$ decreases, $\dot{M}$ decreases correspondingly. As a result, these two effects approximately cancel, leaving the truncation radius $R_T$ nearly unchanged.

What appears to be more relevant for determining the truncation radius is the stellar rotation rate. The models with the same 12-day stellar rotation period exhibit similar truncation radii, whereas the NRT model—whose star has an equally strong field as the REF model but the star does not rotate—shows a truncation radius smaller by approximately 15\%. This trend is expected: stellar rotation transfers angular momentum into the inner disk, which acts to push the truncation radius outward. We quantify the contribution of stellar rotation in the following subsection.
\subsection{Contribution of stellar rotation}
\label{sec:stellar_rotation}
The stellar rotation and stellar magnetic field strength not only contribute to the truncation radius but also to the outflow rate. \citet[][]{Tu2026a} show that the star contributes significantly to the energy and angular momentum of the outflow. Here, we quantify the contribution of the star to angular momentum and energy on the outflow and the disk.

\begin{figure*}
    \centering
    \includegraphics[width=0.9\linewidth]{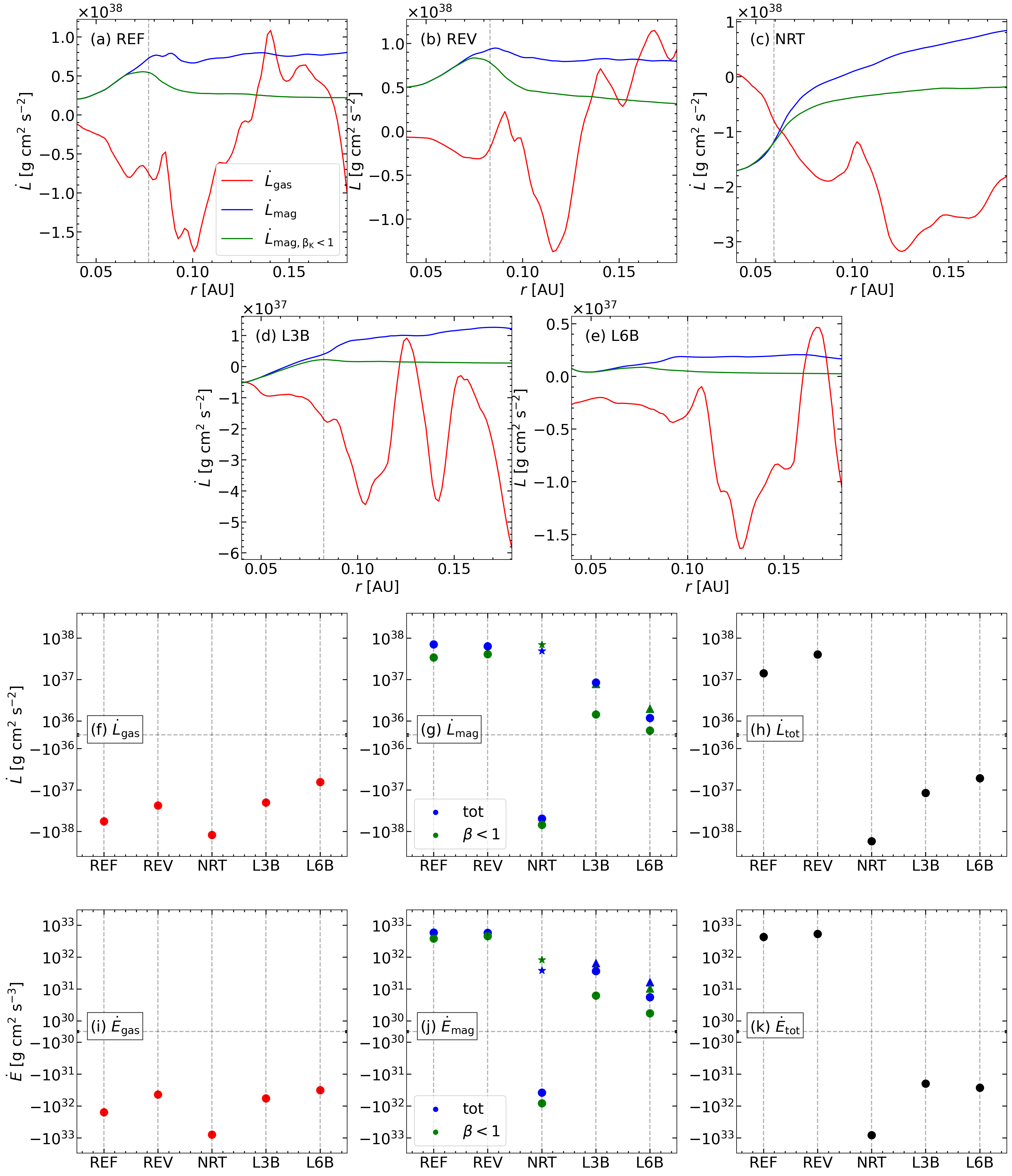}
    \caption{Angular momentum and energy flux around the star in each model. \textbf{Panels (a)-(e)} show the instantaneous angular momentum flux transported by gas ($\dot{L}_\mathrm{gas}$), by magnetic field ($\dot{L}_\mathrm{mag}$), and by magnetic field in the magnetically-dominated region ($\dot{L}_\mathrm{mas, \beta_K<1}$) through spheres of each radius in each model, respectively. \textbf{Panels (f)-(k)} show a quantitative comparison of the fluxes in each model. Each filled circle represents the time-averaged value at $1.3\times$ the truncation radius in each model. \textbf{Panels (f)-(h)} show the angular momentum flux and \textbf{panels (i)-(k)} show the energy flux. The ``star'' symbols in panels (g) and (j) are the negative of the corresponding value in the NRT model. The triangles in panels (g) and (j) indicate the expected values scaled from the REF model (see Section~\ref{sec:stellar_rotation}).}
    \label{fig:torque_on_star}
\end{figure*}

Both angular momentum and energy can be carried by both gas and magnetic field, and we separate them to quantify their respective contributions. To study the stellar contribution, we focus on the spherical-radial ($\hat{r}$) directions. The angular momentum flux carried by the gas through a sphere of radius $r$ is
\begin{equation}
    \dot{L}_\mathrm{gas}=\int\rho v_\phi R~v_rdA,
\end{equation}
where the integration is over the area of the sphere. The angular momentum carried by the magnetic field is given by
\begin{equation}
    \dot{L}_\mathrm{mag}=-\int R\frac{B_\phi B_r}{4\pi} dA,
\end{equation}
which can be derived using arguments similar to those for the classical conserved quantities $l$ \citep[see, e.g.,][]{Ferreira2006, Jacquemin-Ide2021, Tu2025b}. Similarly, the energy fluxes can be derived similarly to the classical conserved $e$ quantity. The gas energy flux is given by
\begin{equation}
    \dot{E}_\mathrm{gas} = \int\frac{1}{2}\rho v^2~v_r dA,
\end{equation}
and energy carried by the magnetic field is
\begin{equation}
    \dot{E}_\mathrm{mag}=\int\Big[\frac{B^2}{4\pi}v_r-\frac{(\mathbfit{v} \cdot \mathbfit{B})B_r}{4\pi}\Big]dA.
\end{equation}
Figure~\ref{fig:torque_on_star} shows these spherical angular momentum and energy fluxes. To further separate the flux into those carried by the low-density plasma (i.e., magnetosphere, disk surface, and outflow), and those by the high-density disk proper, we divide the magnetic angular momentum and energy flux into two components by $\beta_K = 1$, with the $\beta_K < 1$ part tracing the low-density component and $\beta_K>1$ part tracing the high-density disk component.

The first two rows of Figure~\ref{fig:torque_on_star} show the angular momentum flux as a function of spherical-$\hat{r}$ radius at 4.0 yr for different models. The angular momentum flux carried by the magnetic field is positive in the REF, REV, L3B, and L6B models, but is negative in the NRT model, particularly the flux in the low-$\beta_K$ region (the green lines). This is consistent with the finding that the star contributes positively to the angular momentum budget. The dashed vertical line shows the representative radius where we measure and average the fluxes over time (shown in the lower six panels of Figure~\ref{fig:torque_on_star}), and is located at about $1.3\times$ the truncation radius, defined by the $\beta_K$ criteria (see Section~\ref {sec:trunc_radius}), at each time. This radius choice is somewhat arbitrary, but it generally resides where the flux plateaus, and is not too far into the disk that the disk contribution dominates.

The lower six panels show the time-averaged angular momentum fluxes (panels f-h) and energy fluxes (panels i-k) measured at $1.3\times$ the truncation radius between 2.0 and 5.5 yr, measured at the representative radius. Although temporal variability is expected in such a dynamic system, these averages provide a useful measure of the net stellar contribution.

Panel (f) in Figure~\ref{fig:torque_on_star} shows the angular momentum transported by the gas. Because gas angular momentum transport is dominated by accretion, the values are negative, as expected\footnote{We note that the counter-rotating jet (Section~\ref{sec:counter-rotation}) could in principle contribute to negative angular momentum transport. However, as we have verified, the angular momentum carried by this counter-rotating component is negligible compared to the accreting angular momentum flux, primarily because the mass in the (counter-rotating part of the) jet is extremely small relative to the accreted mass.}. Their magnitudes correlate with the accretion rate: largest in the NRT model, smaller in the REF and REV models, and smallest in the L3B and L6B models.

Panel (g) shows the total angular momentum transported by the magnetic field and the portion carried specifically through the low-$\beta_K$ region, respectively. In all rotating-star models (REF, REV, L3B, L6B), the values are positive, indicating magnetic field is transporting angular momentum flux away from the star. The magnetic angular momentum transport rate in the NRT model is negative, indicating that angular momentum is transferred from the disk to the star. This is expected because, in the NRT, the disk rotates faster than the star at all radii. As a result, magnetic field lines connecting the disk and the star transfer angular momentum from the disk to the star. Interestingly, its magnitude is comparable to that in the REF and REV models. Moreover, the amount of angular momentum transported through the low-$\beta_K$ region in the NRT model exceeds the net transport rate, implying that the disk itself is transporting angular momentum outward, a behavior similar to standard $\alpha$-disk accretion models \citep[e.g.,][]{Hartmann1998, Armitage2011}.

The total magnetic angular momentum loss by the star in the REF, REV, L3B, and L6B models decreases sharply with decreasing stellar magnetic field strength, scaling approximately as $B^2$: relative to REF, the loss is lower by factors of $\sim 9$ and $\sim 36$ in the L3B and L6B models, respectively.

The fraction of this angular momentum that enters the low-$\beta_K$ region: i.e., the component that powers the jet—shows a similar decline. In the REF and REV models, roughly 50\% of the magnetic angular momentum flux is directed into the low-$\beta_K$ region, whereas in the L3B and L6B models this fraction drops to $\sim 20\%$. This reduced efficiency leads to a disproportionately weaker jet (the jet mass-loss rate declines faster than $B^2$; see Figure~\ref{fig:overview_acc_out}). A similar trend is also seen in the total energy extraction rate by magnetic field, which scales roughly with $B^2$ (Figure~\ref{fig:torque_on_star}[j]), but not energy into the low-$\beta_K$ region. 

The reduction in angular momentum and energy transportation efficiency also results in the difference in the total angular momentum flux, defined as the sum of gas and magnetic field (i.e., $    \dot{L}_\mathrm{tot}=\dot{L}_\mathrm{gas}+\dot{L}_\mathrm{mag}$; total energy flux is defined similarly). Figure~\ref{fig:torque_on_star}[h] and [k] show the total fluxes in each model. In the models with a rotating star and a strong stellar magnetic field (the REF and REV models), angular momentum and energy are carried away from the star, resulting in a net stellar spin-down. In the NRT model, since the star is non-rotating, both angular momentum and energy are delivered to the star, as expected. In the L3B and L6B models, where the star is spinning but with a lower stellar magnetosphere field strength, the net flux delivers angular momentum and energy to the star, albeit with a magnitude smaller those in the stronger field models, leading to a stellar spin-up.

Since the magnetic field budget is crucial in driving both angular momentum and energy fluxes, in the following subsection, we will show quantitatively that the decrease in magnetic angular momentum and energy transportation rate in the L3B and L6B models is brought by a disproportionate decrease in their available magnetic flux in the low-$\beta_K$ region compared to the models with a stronger stellar magnetic field.

\subsection{Opened stellar magnetosphere flux}
\label{sec:opened_stellar_flux}
We have shown in Section~\ref{sec:mag_flux_evol} that the opened flux in the L3B and L6B models is only about 20\% of that in the REF model. Here, we quantify that this fraction is disproportionately small compared with what one would naively expect.

A simple estimate of the amount of flux that should open can be obtained by integrating the stellar vertical magnetic flux threading the disk outside the truncation radius, since these field lines experience strong differential rotation relative to the star and are therefore prone to opening. This leads to the estimate
\begin{equation}
    \Phi_{\rm open}
    = \int_{R=R_T}^{\infty} B_{z,\star}(R, z=0)  dA ,
    \label{equ:phi_open}
\end{equation}
where $R_T=R_T(M_\star, r_\star,B_\star,\dot{M})$ is the truncation radius.
For an aligned stellar dipole interacting with the disk midplane, the vertical field component is $B_{z,\star}(R,z=0) \propto \bar{\mathbf{m}} R^{-3},$
(see Sec.~\ref{sec:method}). Substituting this into Eq.~\ref{equ:phi_open} yields
\begin{equation}
\Phi_\mathrm{open} \propto \frac{  \bar{\mathbfit{m}}}{R_T}.
\end{equation}

We have shown in Section~\ref{sec:trunc_radius} that the truncation radius $R_T$ depends primarily on the stellar mass. Using the usual scaling for $R_T\propto M_\star^{-1/7}$, we can rewrite the open flux as
\begin{equation}
    \Phi_{\rm open} \propto \bar{\mathbfit{m}} M_\star^{1/7}.
\end{equation}
Thus, in this naive picture, the amount of stellar magnetospheric flux that should open is linearly proportional to the stellar magnetic moment (or equivalently, to the surface magnetic field strength). However, the actual amount of opened magnetospheric flux in the L3B and L6B models is significantly smaller than this estimate. Equivalently, a disproportionately large fraction of the stellar magnetosphere remains closed in these models.

One contributing factor to the enhanced closed-flux fraction is the presence of a pre-existing large-scale magnetic field in the outer disk, which becomes comparatively more important when the stellar magnetosphere is weaker. Although the open stellar magnetosphere dominates the magnetic flux in the inner, jet-launching region, the outer disk field exerts an inward-directed magnetic pressure that opposes the opening of additional stellar field lines and promotes a more dipolar field geometry (see, e.g., the field lines in Figure~\ref{fig:mass_loading}). As a result, this external magnetic support suppresses both the expansion and the partial opening of the magnetosphere.

Because fewer stellar field lines open, fewer ``two-legged'' field lines are available to participate in jet launching. Consequently, the jets in the L3B and L6B models are disproportionately weaker, reflecting the limited supply of open stellar magnetic flux feeding the jet base in these two models. 

The reduction in opened stellar flux not only weakens the jet but can also allow the surrounding disk wind to overpower it, leading to a one-sided jet or even a complete absence of jet signatures at large altitudes. We examine this consequence in detail in the next subsection.

\subsection{Stability of the jet-disk wind system}
\label{sec:one-sided jet}
One striking feature of the NRT, L3B, and L6B models is that the jet can become one-sided—or even disappear entirely—at large heights. For example, in Figure \ref{fig:overview_2D}, both the REF and REV models maintain a strong bipolar jet in the shown snapshot. The NRT model also shows a bipolar jet at this moment, although earlier in the evolution ($t=3.5$ yr) its jet was temporarily choked. In contrast, the L3B and L6B models lack a coherent jet at large distances ($z\geq 6$ au) at the displayed time. At later times, the L6B model loses its large-scale jet entirely, although a small-scale jet persists closer to the disk. For a more complete picture of how the jet polarity evolves over time in each model, we encourage readers to view the animated version of Figure \ref{fig:overview_2D} (a link is provided in its caption).

The main reason why the jet becomes absent or only transient in the NRT, L3B, and L6B models is that it is choked by the disk wind. As seen in the animated version of Figure~\ref{fig:overview_2D}, a jet is indeed launched in both the L3B and L6B models, but it fails to propagate to large distances: it encounters dense disk-wind material, rapidly decelerates, and stalls. As a result, only the disk wind remains visible at large heights. In contrast, the REF and REV models maintain a strong, persistent, and bipolar jet throughout the evolution. This discrepancy raises an important question: why is the jet choked by the disk wind—becoming one-sided or entirely absent—in the NRT, L3B, and L6B models, but not in the REF or REV models?

\begin{figure}
    \centering
    \includegraphics[width=\linewidth]{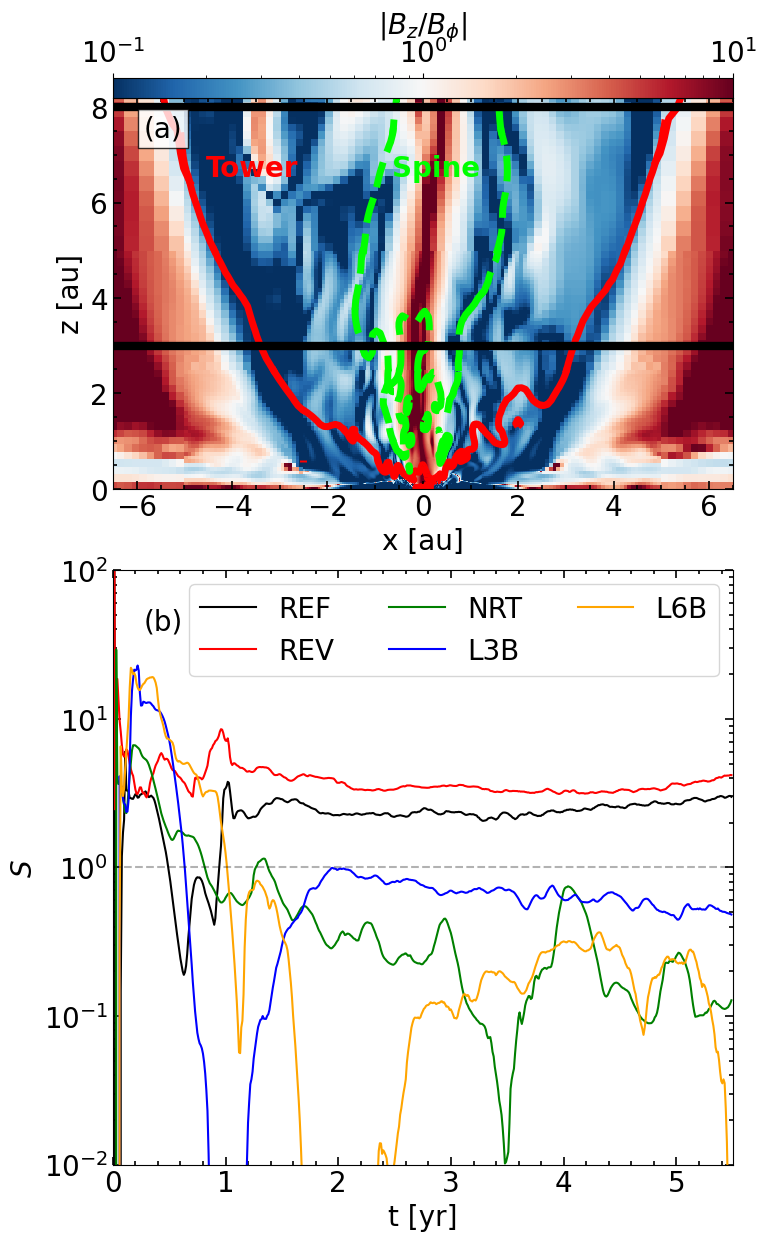}
    \caption{\textbf{Panel (a):} illustration of the ``spine-Tower'' structure with the NRT model. The green contour marks the boundary between the ``spine'', which is defined as the jet itself, and the ``tower'', defined as the disk wind (see section~\ref{sec:one-sided jet}); the red contour marks the boundary of the disk wind ``tower'' and the surrounding disk atmosphere. \textbf{Panel (b)} shows the evolution of the stability criterion $S$ (equation~\ref{equ:spine-tower-stable}) between 3 au and 8 au (between the horizontal black lines in panel [a]) over time in each model.}
    \label{fig:energy_flux_z}
\end{figure}

\citet[]{Tu2025b, Tu2026a} show that both the jet and the surrounding disk wind are driven by toroidal magnetic pressure. The disk wind, in particular, carries a substantial toroidal-field component, so that at high altitudes the magnetic geometry consists of an axial, poloidal-field-dominated jet spine encased by a toroidal-field-dominated loop tower formed by the disk wind, as illustrated in Figure~\ref{fig:energy_flux_z}(a). Such a loop-tower configuration is naturally susceptible to kink instability due to its strong toroidal component \citep[e.g.,][]{Moll2008, Huarte-Espinosa2012}\footnote{ Another potential source of instability is the Kelvin–Helmholtz instability (KHI), which arises from the velocity shear between the fast-moving jet and the more slowly moving surrounding disk wind. However, we do not observe significant KHI growth in our simulations. Moreover, when the magnetic field is strong and predominantly toroidal—as is the case in our models—the kink instability is generally expected to grow more rapidly than the KHI \citep[][]{Hardee2011} and therefore dominate the dynamical evolution of the outflow. We therefore expect the KHI to play only a minor role in shaping the overall jet structure.}. The (axial) poloidal field in the jet acts as a magnetic ``spine,'' providing a stabilizing tension that suppresses the growth of kink modes. This stabilization works only as long as the poloidal field remains sufficiently strong relative to the surrounding toroidal field tower—a situation directly analogous to the Kruskal-Shafranov \citep[K-S,][]{Kruskal1958, Shafranov1956} criterion in tokamak physics, where a dominant toroidal field leads to a destabilizing kink that overwhelms the restoring poloidal tension. In the context of astrophysical jets, such instability is typically categorized as a ``Current-driven Instability'' \citep[CDI, e.g.,][]{Begelman1998, Hardee2011}.


There is, however, one important aspect of the outflow structure that is not captured by the purely magnetic ``spine--tower'' picture described above: the contribution of gas kinetic energy along the spine. In most previous studies, the flow velocity is assumed to be relatively uniform across the spine--tower structure. Under this assumption, the spine and tower effectively co-move with the gas, or equivalently, the gas can be treated as nearly stationary in a co-moving frame. Instabilities are then largely confined to the magnetic structure itself, typically producing internal flow substructure while leaving the overall outflow intact \citep[][]{Guan2014, Dong2020}.

In our simulations, however, the jet flowing along the spine moves substantially faster than the surrounding disk wind that forms the tower, making the co-moving picture inapplicable. At large heights, the spine becomes kinematically dominated, implying that the kinetic energy of the jet provides an additional source of resistance against deformation of the surrounding toroidal-field-dominated tower. Physically, the onset and growth of the kink instability are governed by the MHD momentum equation (Equation~\ref{equ:mhd momentum equation}). Classical linear analyses \citep[e.g.,][]{Begelman1998, Appl2000, Baty2005, Bonanno2011} and most numerical studies \citep[e.g.,][]{Lery2000, Mizuno2009, Narayan2009, Mizuno2012, ONeill2012, Porth2013} typically assume that the inertial term, $(\mathbfit{v}\cdot\nabla)\mathbfit{v}$, is small compared to the magnetic force. This approximation is reasonable when the spine and tower move at comparable speeds, and a co-moving frame can be adopted. In our simulations, however, the spine (jet) is moving much faster than the tower (disk wind), making the inertial term dynamically important.

Since the kink instability originates from a force imbalance in the cylindrical radial direction, the momentum equation may be approximated as
\begin{equation}
    \frac{dv_R}{dt}=v_z\frac{dv_R}{dz}+\frac{1}{4\pi}(\mathbfit{J}\times\mathbfit{B}),
    \label{equ:kink_instability_momentum_equ}
\end{equation}
where gravitational and pressure forces have been neglected because they are small compared to the inertial and magnetic forces. We further assume that, prior to the onset of instability, $v_R\ll v_z$ and $v_\phi\ll v_z$. Equation~\ref{equ:kink_instability_momentum_equ} closely resembles Equation 2.2 of \citet[][]{Begelman1998}, except that we retain the inertial term, neglect the pressure-gradient term, and keep the magnetic force in its full form. 
A key distinction from previous analyses is that the axial velocity is highly non-uniform across the outflow. To leading order, in the cylindrical-$\hat{R}$ direction, it may be approximated as
\begin{equation}
    v_{z,~\mathrm{spine}}\approx v_{\rm jet},
    \quad\quad
    v_{z,~\mathrm{tower}}\approx 0,
\end{equation}
reflecting the large velocity contrast between the jet spine and the surrounding disk wind. Consequently, any kink deformation of the tower must also bend the fast-moving spine enclosed within it. As the instability develops, the spine is displaced radially, causing the gradient term $dv_R/dz$ near the spine-tower boundary to grow rapidly. If the instability continues to develop, this term can approach $\sim v_z/L$, where $L$ is the characteristic growth length scale. In effect, the fast-moving spine introduces an additional inertial restoring force that is absent in the standard co-moving treatment. This is equivalent to adding a gas inertia-force term in the stability criteria derivation in, e.g., \citet[][]{Appl2000}.



{Motivated by the physical insight above, and the fact that the first and second terms of equation~\ref{equ:kink_instability_momentum_equ} can also be interprated as gradients of kinetic and magnetic energy respectively,} we quantify the competition between the stabilizing spine and the destabilizing toroidal tower by defining the spine's stabilization power as the kinetic and magnetic energy associated with the jet ($v_z^p > 10^7~\mathrm{cm~s^{-1}}$); and the destabilization power of the tower is taken to be the toroidal magnetic energy in the surrounding disk wind and jet region ($v_z^p > 10^6~\mathrm{cm~s^{-1}}$, $|B_z/B_\phi|<1$). We then introduce a dimensionless stability parameter,
\begin{equation}
    S \equiv \frac{E_{\mathrm{z,~kin,~jet}} + E_{\mathrm{z,~mag,~jet}}}
    {E_{\mathrm{\phi,~mag,~dw}}}
    =
    \frac{\int_{\rm jet}\left( \rho v_z^2 + \frac{B_z^2}{4\pi} \right) dV}
    {\int_{\rm dw+jet} \frac{B_\phi^2}{4\pi} dV},
    \label{equ:spine-tower-stable}
\end{equation}
which provides a measure of the relative strength of the axial (stabilizing) and toroidal (destabilizing) components. In the magnetically dominated limit, this definition carries essentially the same physical meaning as the Kruskal-Shafranov criterion: both characterize the ratio of poloidal to toroidal field strength that governs kink stability; in the kinematically dominated limit, this definition shares similar physical meaning with the kinetic-$\beta$ as the energy ratios between kinetic energy and magnetic energy.

Figure~\ref{fig:energy_flux_z}(b) shows the time evolution of the azimuthally and vertically averaged $S$ between 3 and 8 au in height for all five models. At late times, only the REF and REV simulations maintain $S>1$, indicating that the axial spine (kinetic plus poloidal magnetic energy) is strong enough to stabilize the toroidal loop tower. In contrast, the NRT, L3B, and L6B cases consistently exhibit $S<1$, consistent with their jets being choked or disrupted by the surrounding disk wind. Empirically, these results suggest that a spine-tower configuration remains stable when $S\gtrsim1$.

{Physically, from a wave-propagation perspective, the dynamical response timescale of the spine is determined by the speed at which information propagates along a jet. This speed is approximately the sum of the jet bulk velocity, $v_\mathrm{jet}$, and the Alfvén speed within the jet, $v_\mathrm{A,~jet}$, giving
\begin{equation}
    t_\mathrm{spine}=\frac{R}{v_\mathrm{spine}}\approx\frac{R}{v_\mathrm{jet}+v_\mathrm{A,~jet}}.    
\end{equation}
Since the jet becomes super-Alfvénic at high altitudes (Figure~\ref{fig:overview_2D}), the kinematic contribution can be comparable to, or even dominate over, the Alfvénic contribution in determining the spine response time.

In contrast, the growth timescale of the kink instability is governed primarily by conditions within the surrounding tower, where the magnetic field is dominated by its toroidal component. The corresponding timescale can therefore be estimated as \citep[][]{Begelman1998, Appl2000, Mizuno2009, Mizuno2012}
\begin{equation}
    t_\mathrm{tower} = \frac{R}{v_\mathrm{A,~tower}}.
\end{equation}
If perturbations in the spine can respond to and counteract the deformation of the tower before the kink instability significantly develops, i.e., if $t_\mathrm{spine}<t_\mathrm{tower}$, the overall spine--tower structure is expected to remain stable. This condition is equivalent to the criterion $S>1$ derived in Equation~\ref{equ:spine-tower-stable}, providing additional physical justification for its applicability in the regime where the spine is kinematically dominated \citep[see also][for a similar idea but in the highly relativistic case]{Bromberg2016}.

Our conclusion is broadly consistent with that of \citet{Tchekhovskoy2016}, who also found that jet stability depends on jet power. Although their study focused on relativistic jets on much larger scales, the underlying connection between jet energetics and stability appears to be qualitatively similar. Future high-resolution simulations that track the jet's long-term propagation will be required to further test and quantify this proposed stability mechanism.}

This empirical stability criterion may provide a means to infer the toroidal magnetic field in the disk wind from the kinematics of protostellar jets. We discuss this implication further in Section~\ref{sec:est_field_ratio}.
\subsection{Counter-rotation in jet}
\label{sec:counter-rotation}
\begin{figure}
    \centering
    \includegraphics[width=\linewidth]{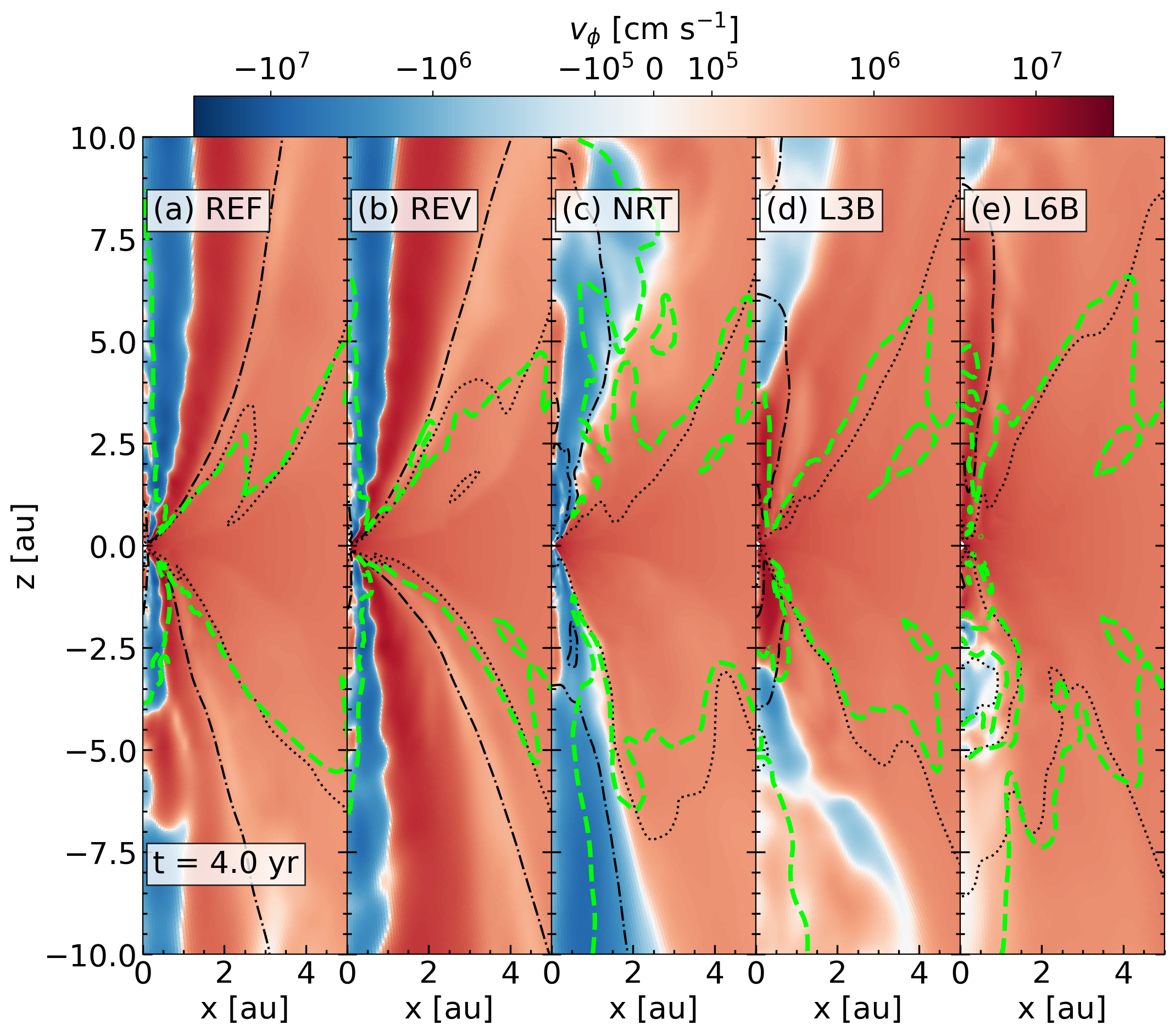}
    \caption{Azimuthally averaged azimuthal velocity $v_\phi$ in each model. Red colors indicate prograde rotation, while blue colors indicate counter-rotation. The dashed and dotted black contours correspond to $v_z^p = 10^7$ and $10^6~\mathrm{cm~s^{-1}}$, respectively, delineating the approximate locations of the outflows. The dashed green contour shows the alfv\'en surface, showing no correlation between this surface and the counter-rotating zones.}
    \label{fig:vphi_compare}
\end{figure}

One notable feature across our simulation suite is that the jet can exhibit partial counter-rotation relative to the disk. Figure~\ref{fig:vphi_compare} shows the azimuthally averaged azimuthal velocity, $v_\phi$, in each model, overlaid with velocity contours to indicate the jet region. In the REF and REV models, the jet displays a two-layer structure: an inner, counter-rotating component near the polar axis, and an outer, co-rotating component that dominates the jet's mass, energy, and angular momentum transport. In the NRT model, the polar region is similarly counter-rotating; however, because the jet is weaker and more narrowly confined, the jet, when present, is dominated by counter-rotating material. Comparable counter-rotating zones are also found in the L3B and L6B models, though they are less pronounced and primarily confined to the lower hemisphere. While some of the apparent counter-rotation arises from geometric projection effects—particularly when the jet is offset from the axis (e.g., the lower hemispheres of the L3B and L6B models in Figure~\ref{fig:overview_2D})—the counter-rotation seen in the REF, REV, and NRT models, where the jet remains closely aligned with the polar axis, more faithfully reflects the intrinsic gas motion.

\begin{figure}
    \centering
    \includegraphics[width=\linewidth]{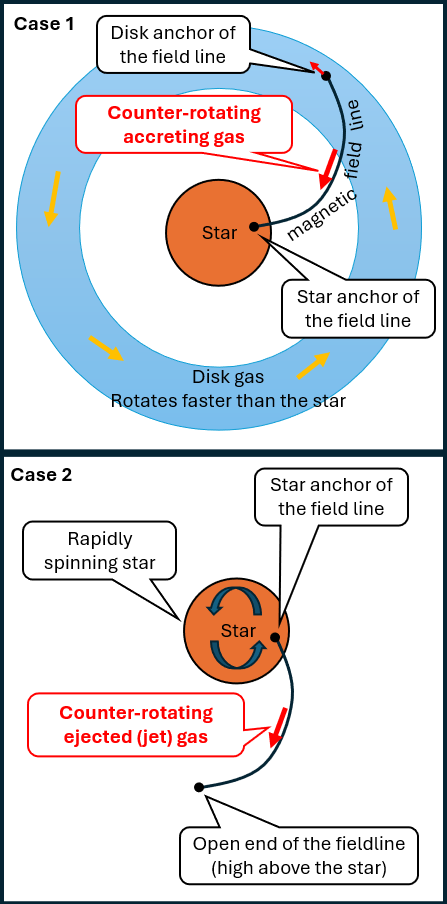}
    \caption{Illustration of two scenarios in which counter-rotation (as seen in Figure~\ref{fig:vphi_compare}) can arise in jets. Both panels are face-on view of the system. \textbf{Upper panel (case 1)}: A slowly rotating star surrounded by a disk, as in the NRT model. Gas streams along a two-legged magnetic field line toward the star and exhibits counter-rotation as a result of differential rotation between the star and the disk. \textbf{Lower panel (case 2)}: A rapidly rotating star threaded by open magnetic field lines, as in the REF and REV models. In this case, the footpoint of the magnetic field line anchored on the star rotates faster than the open end of the field line. Gas loaded onto the field line and accelerated outward, therefore exhibits counter-rotating motion.}
    \label{fig:sketch_counterrotation}
\end{figure}

At first glance, the presence of counter-rotating gas may appear surprising, since the rest of the system, including the star and the disk proper, rotates in a single, coherent direction. However, counter-rotation can arise naturally from the flow along magnetic field lines that are deformed by differential rotation between the gas and the star. There are two generic ways in which this occurs, illustrated schematically in Figure~\ref{fig:sketch_counterrotation}, assuming ideal MHD so that the gas motion is tied to the magnetic field lines. In the first case, the gas rotates faster than the star; gas accreting along a field line toward the star then appears counter-rotating in the inertial frame \citep[similar to the mechanism in][see also \citealp{Zhu2024}]{Ghosh1979}. In the second case, the star rotates faster than the gas at high altitudes; material flowing outward along these star-anchored, open field lines is driven into counter-rotation by the stellar twisting of the field. This counter-rotation is typically more pronounced after the jet is launched and at higher altitudes, where the gas propagates ballistically along the field lines.\footnote{We note that the appearance of counter-rotation at high altitudes can also arise from gas advection, where counter-rotating material at low altitudes is transported upward by the jet flow. In addition, localized islands of counter-rotating or forward-rotating gas embedded within an otherwise coherently rotating outflow may result from magnetic responses to over-twisting, which can induce localized azimuthal motions. These effects, however, are secondary and comparatively subtle; the dominant drivers of counter-rotation in our simulations remain the two mechanisms illustrated in Figure~\ref{fig:sketch_counterrotation}.}

The first mechanism likely dominates in the NRT model, where the disk always rotates faster than the non-rotating star. As a result, any gas approaching the star along magnetic field lines exhibits counter-rotation. Since the jet is launched along the ``two-legged'' field lines, it is likely counter-rotating at launch and maintains this sense of rotation as it propagates to larger heights. A similar process can also operate, to a more limited extent, in the REF and REV models within the region between the truncation radius and the corotation radius, where the disk rotation exceeds that of the star. However, this region is relatively narrow compared to the NRT case. As a result, in these models, the second mechanism, acting after the outflow is launched, can become increasingly important at higher altitudes. In particular, gas loaded onto star-twisted polar field lines can be driven into counter-rotation, especially near the axis where the magnetic spine is strongest.

It is important to note that our mechanism differs from the analytic picture of \citet{Sauty2012}, in which counter-rotation arises from energy transfer between the magnetic field and the gas near the Alfv\'en surface. In our case, the counter-rotating component shows no clear correlation with the Alfv\'en surface (green contours in Figure~\ref{fig:vphi_compare}). Moreover, such counter-rotation is not reported in \citet{Zhu2025}, despite their use of a similar magnetospheric treatment but with a smaller computational domain. Given that the counter-rotation in our L3B and L6B models appears primarily at high altitudes, this discrepancy is likely due to their weaker stellar magnetic field strength and more limited spatial extent, which may prevent the development or capture of this feature at high altitudes. We also note that counter-rotation is generally not expected in classical self-similar jet solutions \citep[see, e.g.,][]{Anderson2003, Ferreira2006}, where the Bernoulli invariants are conserved along each magnetic field line in the jet. While these conserved integrals can provide useful physical insight in steady-state configurations, they are not generally applicable in our highly dynamic jet-launching scenario, which involves continuous changes in magnetic topology driven by localized magnetic reconnection \citep[see further discussion in][]{Tu2025b, Tu2026a}.

Although the counter-rotating component contributes only modestly to the total jet in the REF and REV models, it constitutes the dominant fast-moving component in the NRT model. The rotational properties of jets—and their connection to stellar spin—therefore provide a potentially powerful observational diagnostic of stellar rotation rates. We discuss this implication further in Section~\ref{sec:poloidal vel-rotation}.

\section{Discussion}
\label{sec:discussion} 
\subsection{Effective star-disk interaction through fieldlines along the disk surface}
\citet[][]{Tu2026a} analyzed the REF model in detail and showed that the jet is launched along ``two-legged'' magnetic field lines that skim the disk surface. In Section \ref{sec:stellar_rotation}, we demonstrate that this mechanism persists across all models with different stellar rotation rates and magnetic field strengths, and that the high-speed jet material consistently originates from the disk surface—fully consistent with the findings of \citet[][]{Tu2026a}.

Outflows driven by magnetosphere-disk interactions have also been identified in previous work, including the ``magnetospheric ejection'' (ME) picture of \citet{Zanni2013} \citep[see also, e.g.,][]{Lynden-Bell1994, Ireland2021, Zhu2025} and the ``propeller'' mechanism studied by \citet{Romanova2005} \citep[see also, e.g.,][]{Romanova2009, Romanova2014}. In \citet{Zanni2013}'s picture, three types of outflows are produced: (1) a stellar wind injected numerically from the inner boundary; (2) the ``magneto-spherical ejection'' (ME), a conical outflow driven by stellar magnetosphere-disk interaction, and (3) the disk wind, launched by open magnetic field lines threading the disk. In \citet{Romanova2005}'s picture, a rapidly rotating star can drive a strong ``propeller'' outflow along stellar magnetosphere field lines if the corotation radius is smaller than the truncation radius. Similar to \citet{Zanni2013}'s ME, a conical-shaped outflow is also identified in \citet{Romanova2005}. 

In \citet{Romanova2005}'s terminology, our jet in the REF, REV, L3B, and L6B is effectively propeller-driven, even though our corotation radius $R_\mathrm{co}\approx0.1~\mathrm{au}$ lies beyond the truncation radius $R_T=0.08~\mathrm{au}$. This is because the stellar magnetospheric fieldlines connect to the elevated disk surface at locations that extend far beyond the corotation radius, so a similar ``propeller'' effect can be active. In the NRT model, the corotational radius is located infinitely far away, yet a jet is still launched because the disk rotation alone provides the required energy to the jet. Using \citet{Zanni2013} terminology, our simulations contain no stellar wind (because we do not impose one at the inner boundary). All outflow is generated self-consistently within the computational domain, leaving only the ME and disk-wind components in their classification. The latter maps directly onto our disk wind.


The jets in our simulations are driven by the magnetosphere-disk interaction. In this sense, they are similar to the ME defined in \citet{Zanni2013}, which is also driven by the interaction, even though the two differ in several crucial aspects,  as discussed in detail in \citet{Tu2026a}. Still, it is instructive to compare the torque exerted on the star by our jet with the ME torque estimated in previous analytic work in \citet{Ireland2021}.

The torque we calculated in Section~\ref{sec:stellar_rotation}, as the ``angular momentum flux transported by magnetic field'', is substantially larger (by one to two orders of magnitude) in magnitude than the estimation using the formalism in \citet{Ireland2021}. The main reason for the torque to be so much more effective in our model than in theirs is that the stellar magnetosphere and the disk interact mostly along the disk surface, so that the radius of interaction $R_\mathrm{int}$, assumed to be at $R_T$ (the truncation radius), becomes at least a few times larger than $R_T$. As a result, the star can transfer angular momentum to the disk even if the truncation radius is smaller than the corotational radius (i.e., $R_T <R_\mathrm{co}$), which is not the case in \citet{Ireland2021}'s formalism. 

Since the formalism in \citet{Ireland2021} does not take into account the stellar spin, we use and modify this formalism to show that a variation of their derivation can be used to provide an order-of-magnitude estimate for the magnitude of star-disk interaction torque. The angular momentum flux at cylindrical radius $R$ is given by
\begin{equation}
    d\dot{L} = q(R)B_\theta(R)^2R^2dR,
    \label{equ:Ireland_eq1}
\end{equation}
where $q(R)$ and $B_\theta(R)$ are the magnetic twist and field strength at $R$, respectively. Following \citet{Ireland2021}, we have
\begin{equation}
    q(R)=K_i\Big[\Big(\frac{R}{R_\mathrm{co}}\Big)^{3/2}-1\Big],
    \label{equ:Ireland_eq2}
\end{equation}
and
\begin{equation}
    B_\theta(R)= B_\star\Big(\frac{R}{R_\star}\Big)^{m_\mathrm{B}},
    \label{equ:Ireland_eq3}
\end{equation}
where $K_i=0.00772$ and $m_B=-2.54$ following \citet{Ireland2021}. Note that the initial magnetic field strength goes as $B_\theta(t=0)\propto R^{-3}$ in our simulation, but taking into account the change in magnetic field geometry due to partial opening of the magnetosphere, $m_B=-2.54$ can be a reasonable estimation. Integrating equation~\ref{equ:Ireland_eq1} using equation~\ref{equ:Ireland_eq2} and \ref{equ:Ireland_eq3} over the interaction zone--i.e., between the truncation radius and a rough outer boundary of effective interaction $R_\mathrm{int}$, we get
\begin{equation}
    \dot{L}\equiv\int_{R_t}^{R_\mathrm{int}}d\dot{L}=\frac{K_BB_\star}{R_\star^{m_B}R_\mathrm{co}^{3/2}}\Big[\frac{R^{p}}{p}-\frac{R_\mathrm{co}^{3/2}R^{q}}{q}\Big]_{R_t}^{R_\mathrm{int}},
\end{equation}
where $p=4.5+m_B$ and $q=3+m_B$ come from the integration; $B_\star$ and $R_\star$ are the magnetic field and radius of the star, respectively. 

Taking our simulation parameters $B_\star=2000~\mathrm{G}$, $R_\star=0.015~\mathrm{au}$, and an interaction radius $R_\mathrm{int}=0.3~\mathrm{au}$, we get $\dot{L}\sim10^{38}~\mathrm{g~cm^2~s^{-2}}$, on a the same order of magnitude as our simulation results. Although the direction of the torque (spin-up or spin-down) depends on the stellar rotation rate, this estimate provides a useful measure of the torque magnitude exerted by the star in launching a jet from the extended disk surface.

The torque we calculated can also be compared with those in \citet{Zhu2025}, who normalize the torque using
\begin{equation}
    n\equiv\frac{\dot{L}}{\dot{M}(GM_\star R_T)^{1/2}}.
\end{equation}
The denominator, evaluated using our simulation parameters in the REF model, scales as
\begin{equation}
\begin{split}
    \dot{M}(GM_\star R_T)^{1/2} & \approx 8\times10^{37}~\mathrm{g~cm^2~s^{-2}}\\
    &\Big(\frac{\dot{M}}{10^{-7}M_\odot~\mathrm{yr}^{-1}}\Big)\Big(\frac{M_\star}{M_\odot}\Big)^{-1/2}\Big(\frac{R_T}{0.08~\mathrm{au}}\Big)^{-1/2},
\end{split}
\end{equation}
implying $n \gtrsim 1$ generally in our suite of models, comparable to the values reported by \citet{Zhu2025}, who present a suite of models with different rotation rates. This agreement, in contrast to the analytic prescription of \citet{Ireland2021} discussed above, further supports the view that magnetospheric interactions extend beyond the truncation radius, and that magnetosphere-disk surface coupling plays a key role in regulating the dynamics and angular momentum exchange.


In summary, both the analytic estimate and the numerical results demonstrate that confining the star-disk interaction solely to the truncation radius severely underestimates the torque. Incorporating the extended disk-surface interaction along the ``two-legged'' field lines identified in \citet[][]{Tu2026a} is therefore essential for capturing the true influence of the stellar magnetosphere on angular momentum transport and outflow launching.

\subsection{Outflow morphology as a probe of toroidal field strength in the disk-wind}
\label{sec:est_field_ratio}
We have shown in Section~\ref{sec:one-sided jet} that the stability of the jet structure depends on the ratio between the total poloidal energy in the jet ``spine'' and the surrounding toroidal energy in the disk wind ``tower''. In this section, we use this stability criterion to propose a method to indirectly estimate the magnetic property using observation of jet speed.

\begin{figure*}
    \centering
    \includegraphics[width=\linewidth]{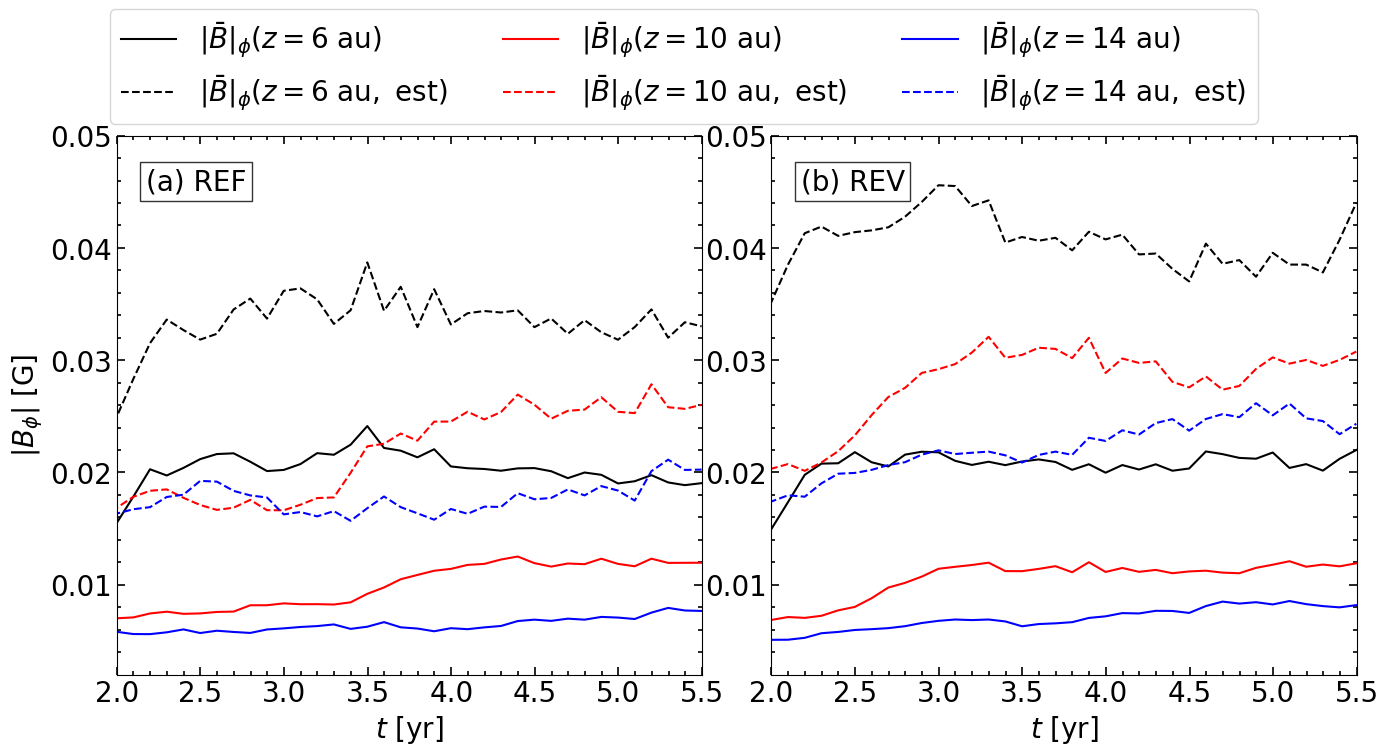}
    \caption{Estimation of the averaged disk wind toroidal magnetic field strength $\bar{B}_\phi$ in the REF (left) and REV (right) models, both of which exhibit a steady jet. The solid lines show the measured averaged toroidal magnetic field strength in the simulation, and the dashed lines show the estimation using the right-hand side of equation~\ref{equ:est_dw_Bphi}, where an overestimation is expected (see section~\ref{sec:est_field_ratio}). The black, red, and blue curves are the measurements and estimations carried out at $z=6, 10$, and 14 au respectively.}
    \label{fig:energy_flux_est_diskBphi}
\end{figure*}

We only consider the case where a stable jet is present and thus observable, so $S\geq1$. Note that the denominator of equation~\ref{equ:spine-tower-stable} is dominated by the kinetic energy term, as the jet is kinetically dominated at large altitudes. So we can write equation~\ref{equ:spine-tower-stable} as
\begin{equation}
    E_\mathrm{B,~\phi,~dw+jet} \leq E_\mathrm{kin,~z,~jet},
    \label{equ:Ejet_to_Edwphi}
\end{equation}
where $E_\mathrm{B,~\phi,~dw+jet}$ is the denominator in equation~\ref{equ:spine-tower-stable} and $E_\mathrm{kin,~z,~jet}$ the numerator, keeping only the kinetic energy term.  Notice that both terms integrate over the same height, so if the energy density is about uniform within the volume, we can write equation~\ref{equ:Ejet_to_Edwphi} as
\begin{equation}
    \mathcal{E}_\mathrm{B,~\phi,~dw+jet}A_\mathrm{dw+jet} \leq \mathcal{E}_\mathrm{kin,~z,~jet} A_\mathrm{jet},
    \label{equ:Evol_to_Edens}
\end{equation}
where $\mathcal{E}$ stands for energy density per unit volume; $A_\mathrm{dw+jet}$ and $A_\mathrm{jet}$ are the horizontal cross-sectional area of the jet and disk wind, and the jet itself, respectively. The estimated upper limit of the azimuthal magnetic field strength of the disk wind is thus given by
\begin{equation}
    |\bar{B}_{\phi,~\mathrm{dw},~\mathrm{est}}|\leq\sqrt{\Big|\frac{4\pi\mathcal{E}_\mathrm{kin,~z,~jet}A_\mathrm{jet}}{A_\mathrm{dw+jet}}\Big|}.
    \label{equ:est_dw_Bphi}
\end{equation}
Figure~\ref{fig:energy_flux_est_diskBphi} shows the averaged azimuthal magnetic field strength in the disk wind and the estimated value in the REF and REV models. As expected, equation~\ref{equ:est_dw_Bphi} yielded an over-estimation, and because of geometric dilution of the toroidal magnetic field, the amount over-estimated increases as $z$ increases. Empirically, using the simulation data, we estimated that the amount overestimated is $\sim(z/\mathrm{AU})^{0.4}$, which extrapolates to about $10\times$ over-estimation at an observable height of $z=300~\mathrm{au}$. 

In practice, a more directly relevant quantity is the poloidal magnetic field strength, $B_z$, in the jet. This can be estimated by combining the inferred toroidal field strength, $B_\phi$, with the observationally determined inclination angle of the helical magnetic field \citep[e.g.,][]{Lee2018, Rodriguez-Kamenetzky2025}. Assuming magnetic flux conservation along the disk wind, the poloidal field strength at the wind-launching region can then be obtained by extrapolating the measured jet field back to its footpoint in the disk.

\subsection{Probing stellar rotation, instead of launching location, with jet rotation}
\label{sec:poloidal vel-rotation}
In connecting observed jet properties with their launching locations, an often-used approach is to employ the rotation--poloidal velocity relation to estimate the jet-launching radius, as originally proposed by \citet{Anderson2003} and refined by \citet{Ferreira2006}, based on 2D (axisymmetric) steady magnetocentrifugal wind theory. This relation is then subsequently applied in, e.g., \citet{Louvet2018, Ai2024, Lopez-Vazquez2024}. The basic idea behind this method is that, by assuming the conservation of certain quantities along an outflow streamline, one can infer the launching radius of the jet by measuring its poloidal velocity $v_{\mathrm{pol}}$ and specific angular momentum $R v_\phi$. In practice, Figure~2 of \citet{Ferreira2006} is commonly used to relate these observed quantities to an inferred launching radius.

The applicability of this relation, however, was already questioned in the first paper of this series, \citet[][]{Tu2025b}, which studied jet launching in the absence of a stellar magnetosphere (i.e., a disk-only model). \citet[][]{Tu2025b} demonstrated that the rotation-poloidal velocity relation should not be used to infer the launching radius when the fast, lightly mass-loaded jet is driven primarily by magnetic pressure. Here, we show that this conclusion also applies to jets launched by magnetosphere-disk interaction. As a result, the use of the rotation-poloidal velocity relation to estimate jet-launching radii should be treated with great caution.

As shown in Section~\ref{sec:counter-rotation}, the jets in the REF and REV models are dominated by a prograde rotation component, whereas the jet in the NRT model is dominated by counter-rotation. We exclude the L3B and L6B models from the following discussion because their jets are unstable; however, we have verified that the conclusions presented here remain valid whenever a jet shows up in the L3B and L6B models.

\begin{figure*}
    \centering
    \includegraphics[width=\linewidth]{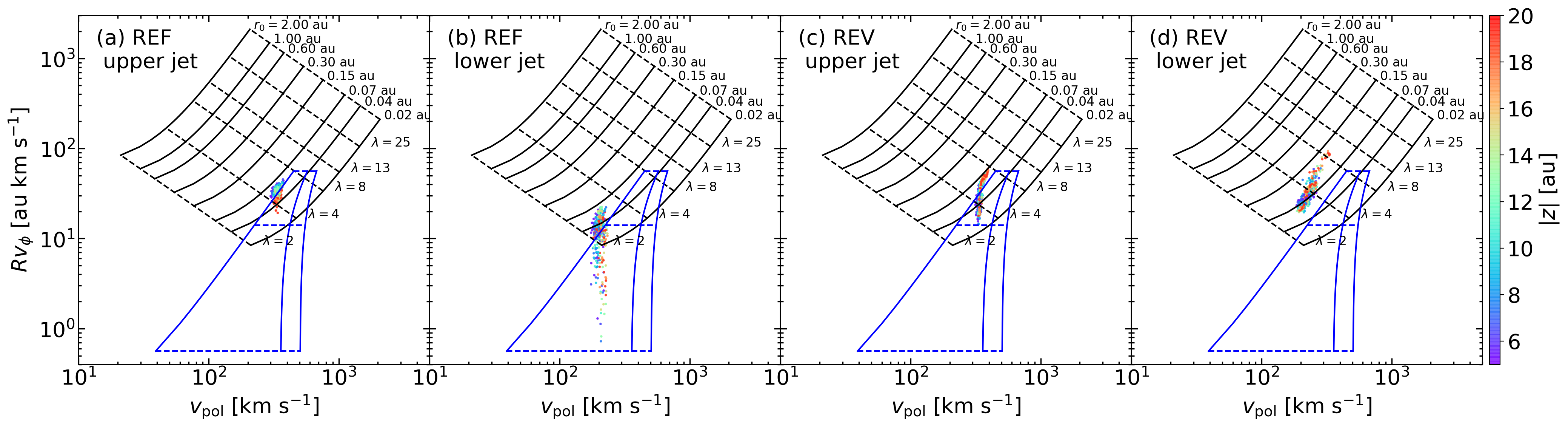}
    \caption{Poloidal velocity-specific angular momentum relation from classical expectations and from our numerical simulations. \textbf{Panels (a) and (b)} show the jet in the upper and lower hemispheres, respectively, in the REF model; \textbf{panels (c) and (d)} show the corresponding results for the REV model. The black and blue solid curves indicate the classical expectations from centrifugally driven and thermally driven wind models, respectively \citep[][]{Ferreira2006}. The jet in our simulations is represented by scatter points, where each point corresponds to a mass-weighted poloidal velocity-angular momentum measurement at a given height and time. In both models, the distributions deviate strongly from the classical predictions and instead exhibit some vertical spread (see Section~\ref{sec:poloidal vel-rotation}).}
    \label{fig:Rvphi_vpol}
\end{figure*}

The rotation-poloidal velocity relation is immediately inapplicable to the NRT model, as counter-rotation is not predicted by the relation. For the REF and REV models, we show in Figure~\ref{fig:Rvphi_vpol} the distribution of the mass-weighted rotation-poloidal velocity relation, measured at different times and heights, with the color of each point indicating the corresponding height. The theoretical prediction in \citet{Ferreira2006} is overplotted for reference. In both models, the distributions do not follow the theoretical predictions. In cases where a trend is present, the distribution exhibits a vertical structure similar to that reported in \citet[][]{Tu2025b}, namely that the rotational velocity can vary largely independently of the poloidal velocity. 

The physical reason why the jets in our models do not follow the predictions of \citet{Ferreira2006} is that their predictions were based on outflows driven magento-centrifugally from a thin disk with the flow rotation speed at large observable distances directly tied to the rotation speed (and thus the radius) at the foot point on the Keplerian disk. This is not the case for our jets, which are driven by the differential rotation between the star and the surface of the magnetically elevated disk. Specifically, they are driven predominantly by the magnetic pressure generated by the twisting of the field lines connecting the star and disk surface, or equivalently by Poynting flux. In this respect, our models share more similarities with the ``thermally driven'' solutions discussed in \citet{Ferreira2006}, since magnetic pressure can be formally written as an effective thermal term, as shown in \citet{Tu2025b}.

As a result, rather than tracing the radius on a Keplerian disk where the jet is assumed to be launched magneto-centrifugally, the rotation-poloidal velocity relation in our models more directly reflects the stellar rotation rate. When the star rotates rapidly, as in the REF and REV models, the jet tends to rotate in the prograde direction. In contrast, when the stellar rotation is slow, as in the limiting case of a non-rotating star represented by the NRT model, the jet can become counter-rotating. This connection with stellar rotation provides a natural explanation for observed counter-rotating jets \citep[e.g., TH28, observed by][]{Coffey2004, Louvet2016}: such systems may indicate a slowly rotating star hosting a relatively strong stellar magnetic field.

\section{Conclusion}
\label{sec:conclusion}
We present a suite of numerical models of the jet-launching inner disk region that self-consistently include magnetosphere-disk interactions. By varying a single physical parameter in each model, we demonstrate that jet properties depend sensitively on the rotational and magnetic properties of the central star. Our main findings are summarized as follows:
\begin{enumerate}

\item The strength and persistence of the jet depend systematically on the stellar magnetospheric properties. Reversing the polarity between the stellar dipole and the disk field does not significantly alter the overall jet power, indicating that the launching mechanism is robust with respect to magnetic alignment or anti-alignment. In contrast, stellar rotation plays a critical role: removing stellar spin substantially reduces the jet power and makes the outflow more susceptible to disruption by the surrounding disk wind. Decreasing the stellar magnetospheric field strength reduces the open stellar magnetic flux, thereby weakening the axial jet and causing it to become intermittent, one-sided, or absent at large heights. Stable, bipolar jets arise only when the stellar magnetic field is sufficiently strong and the star rotates at a significant rate.

\item At large heights, the outflow naturally organizes into a kinetically dominated axial jet, threaded predominantly by a poloidal magnetic field (``spine'') and surrounded by a toroidal-field-dominated disk-wind ``tower.'' The stability of this configuration is governed by the ratio of stabilizing axial energy (kinetic plus poloidal magnetic) to destabilizing toroidal magnetic energy in the surrounding wind. When this ratio exceeds unity, the axial structure remains stable and a bipolar jet persists; when it falls below unity, the jet is choked or disrupted by the disk wind. Weaker stellar magnetospheric fields and the absence of stellar rotation both diminish the axial component, thereby lowering this stability parameter and leading to one-sided or absent jets. This stability criterion provides a natural explanation for the observed diversity in jet morphology and offers a means to constrain the toroidal magnetic field strength in the disk wind surrounding a well-developed bipolar jet using observable quantities.

\item The parameter exploration of stellar magnetospheric field strength and rotation presented here reinforces the picture of accretion-fed, star-anchored Poynting jets in the low-density polar cavity first identified by \citet{Tu2026a}. In particular, we show that the fast jet material observed at large heights along the opened magnetospheric field lines anchored on the star originates near the elevated disk surface, where magnetosphere-disk interaction proceeds through ``two-legged'' field lines.

\item We find that magnetic extraction of stellar spin angular momentum is significantly enhanced by the connection of magnetospheric field lines to the elevated disk surface well beyond the nominal disk truncation radius, with a substantial fraction of the extracted angular momentum carried away by the axial jet. The jet thus plays an important role in the star's spin-down.

\item Counter-rotation is present in all models and is particularly prominent in the non-rotating star and stronger stellar field cases. These counter-rotating regions arise naturally from gas-magnetic field interactions in the jet-launching region. In addition to identifying counter-rotating material, we demonstrate that the commonly used relation between poloidal velocity and specific angular momentum cannot be reliably used to infer the jet-launching radius when jets are driven by magnetosphere-disk interaction, as in our models, rather than magnetocentrifugally from a Keplerian disk alone. In this framework, the relation instead serves as a diagnostic of the stellar rotation rate, since the spatial extent of counter-rotation depends sensitively on stellar spin.

\end{enumerate}
\begin{acknowledgments}
We thank Hsien Shang and Lee Hartmann for insightful and stimulating discussions. YT acknowledges an interdisciplinary fellowship at the University of Virginia. YT is supported by NASA Emerging Worlds 80NSSC24K1285. ZYL is supported in part by NASA 80NSSC20K0533, NSF AST-2307199, JWST-GO-02104.002-A, JWST-GO-08872.003-A, and the Virginia Institute of Theoretical Astronomy (VITA). Z.Z. acknowledges support from
NASA award 80NSSC25K7144 and NSF award 2429732
and 2408207. We acknowledge computing resources from the University of Virginia Research Computing (RC), NASA
High-Performance Computing, the 
EPL HPC at Carnegie Science Earth and Planets Laboratory Division, the computational resources and services provided by Advanced Research Computing at the University of Michigan, Ann Arbor, the Anvil Supercomputer at Purdue University, and Stampede3 Supercomputer at the University of Texas, made available through NSF's ACCESS computing allocation PHY250098 and PHY260035.

\end{acknowledgments}

\software{\texttt{Athena++} \citep[][]{Stone2020}, NumPy \citep{Numpy}, Matplotlib \citep{Matplotlib}, Numba \citep{numba}}







\bibliography{sample7}{}
\bibliographystyle{aasjournalv7}



\end{document}